\documentclass[letterpaper,twocolumn,10pt]{article}
\usepackage{usenix-2020-09}

\usepackage{graphicx}
\graphicspath{{figs/}}

\usepackage[ruled,linesnumbered]{algorithm2e}
\usepackage{amsmath,amsfonts}
\usepackage{xspace}
\usepackage{xcolor}
\usepackage{enumitem}
\usepackage{booktabs}
\usepackage{subcaption}
\usepackage{tcolorbox}
\usepackage{multicol}
\usepackage{xurl}
\SetCommentSty{mycommfont}

\newcommand{\para}[1]{\smallskip\noindent\textbf{#1}~}
\newcommand{\sysname}{Autothrottle\xspace}
\newcommand{\sysglobal}{Tower\xspace}
\newcommand{\syslocal}{Captain\xspace}
\newcommand{\syslocals}{Captains\xspace}
\let\oldtexttt\texttt
\renewcommand\texttt[1]{\oldtexttt{\fontfamily{lmtt}\selectfont\small#1}}

\newcommand\tightsection[1]{\vspace{-5pt}\section{#1}}

\let\oldsubsection\subsection
\renewcommand\subsection[1]{\vspace{-5pt}\oldsubsection{#1}}

\let\oldsubsubsection\subsubsection
\renewcommand\subsubsection[1]{\vspace{-5pt}\oldsubsubsection{#1}}

\usepackage{titling}
\newcommand{\authdag}{\textsuperscript{\textdagger}}
\newcommand{\authsec}{\textsuperscript{\textsection}}
\newcommand{\authp}{\textsuperscript{\textparagraph}}

\begin{document}

\title{\Large \bf {\sysname}: A Practical Bi-Level Approach to \\
Resource Management for SLO-Targeted Microservices \\ \vspace*{0.8em}
  \scalebox{1.12}{\normalfont\normalsize
   Zibo Wang\authdag\authsec,
   Pinghe Li\authp,
   Chieh-Jan Mike Liang\authsec,
   Feng Wu\authdag,
   Francis Y. Yan\authsec}\\ \vspace*{-0.2em}
  {\normalfont\normalsize
   \authdag University of Science and Technology of China,
   \authp ETH Zurich,
   \authsec Microsoft Research} \vspace*{-3.8em}
}

\author{}
\date{}

\maketitle

\begin{abstract}

Achieving resource efficiency while preserving end-user experience is
non-trivial for cloud application operators. As cloud applications progressively
adopt microservices, resource managers are faced with
two distinct levels of system behavior: end-to-end application
latency and per-service resource usage.
Translating between the two levels, however, is challenging because
user requests traverse heterogeneous services that collectively (but unevenly)
contribute to the end-to-end latency. We present \textit{{\sysname}}, a
bi-level resource management framework for
microservices with latency SLOs (service-level objectives).
It architecturally decouples
application SLO feedback from service resource control,
and bridges them through the notion of performance targets.
Specifically, an application-wide learning-based controller is employed
to periodically set performance targets---expressed as CPU
throttle ratios---for per-service heuristic controllers to attain.
We evaluate {\sysname} on three microservice applications,
with workload traces from production scenarios.
Results show superior CPU savings,
up to 26.21\% over the best-performing baseline
and up to 93.84\% over all baselines.
\end{abstract}

\tightsection{Introduction}
\label{sec:intro}

To ensure a seamless end-user experience, many user-facing
latency-sensitive applications impose an SLO (service-level objective) on
the end-to-end latency. Traditionally, cloud application operators resort
to resource over-provisioning to avoid SLO violations, yet doing so unnecessarily
wastes resources~\cite{morpheus, pegasus}.  Previous efforts have demonstrated
significant savings if the excess resources could be harvested or reclaimed for
co-located applications in a multi-tenant environment~\cite{smartharvest,
scavenger, perfiso, heracles, harvested_serverless, harvest_vm}.

A key enabler for such resource saving is SLO-targeted resource management. Its
goal is to continuously minimize the total resources allocated, while still
satisfying the end-to-end latency SLO. Unfortunately, modern cloud applications
can be beyond current resource managers, due to the progressive shift from
monolithic to distributed architecture~\cite{twitter_microservices,
netflix_microservices, uber_microservices, alibaba_microservices,
wechat_overload}. They are a topology of cloud-native services or
microservices\footnote{In this paper, we use ``services'' and ``microservices''
interchangeably.}, and user requests traverse a chain of execution dependencies
among services of logic, databases, and machine learning (ML) model serving.
Notably, this creates distinct levels of system behavior---the macro perspective
reveals the end-to-end performance (e.g., user request latencies) and SLO, and
the micro perspective is scoped to local
measurements (e.g., service CPU usage) and resource control.

\begin{figure}[t]
  \centering
  \includegraphics[width=\columnwidth]{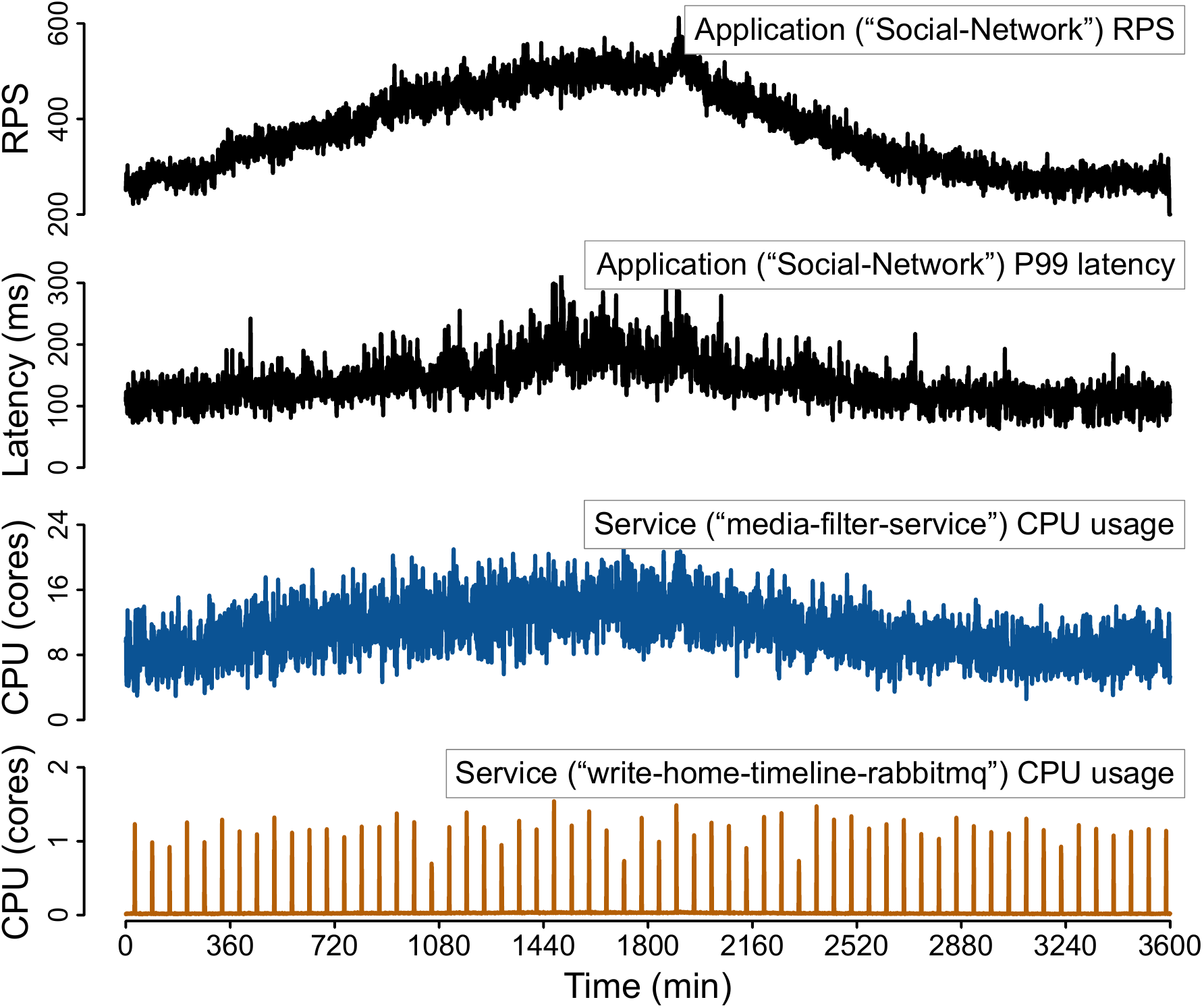}
  \caption{Individual microservices (bottom two panels) can exhibit vastly different resource usage patterns and short-term fluctuations.
  In addition, they do not necessarily have a strong correlation with the end-to-end application-level measurements (top two panels).}
  \label{fig:macro_vs_micro}
  \vspace{-8pt}
\end{figure}

The distributed nature of microservices brings unique implications to resource
management. First, heterogeneous services can exhibit vastly different resource
usage patterns, due to how various user requests stress each service. The
bottom two panels in Figure~\ref{fig:macro_vs_micro} contrast the CPU usage of two
services in an application, Social-Network~\cite{deathstarbench}. Second,
application performance and per-service resource usage are measurements at
different levels, without necessarily exhibiting a strong correlation (top two vs. bottom two panels in Figure~\ref{fig:macro_vs_micro}). Translating
between them requires knowing each user request's actual resource requirements and
its service-to-service execution flow. Moreover, this execution chain incurs
undesirable delays in observing effects of allocation changes on the
end-to-end performance, further complicating resource management.

At first glance, it appears that resource managers could implicitly address
the distributed nature by either considering application-wide
dependencies~\cite{sinan, firm, graf, phpa, alphar} or
employing heuristics with operator-defined rules
on individual services~\cite{k8s_autoscale}.
The former centralizes resource control with a global view of the service
topology, while the latter delegates control to each service that
acts on locally observed resource usage.
Nevertheless, maintaining a global view is susceptible to topology changes and
evolution~\cite{wechat_overload,seshagiri2022sok},
and relying solely on local speculations may not achieve global optimality.

Instead, we embrace the distinct levels of distributed system behavior, and
architecturally decouple mechanisms of application-level SLO feedback
and service-level resource control. We design \textit{{\sysname}}, a
bi-level learning-assisted resource management framework for SLO-targeted
microservices. The goal is to better use the visibility into application
performance and SLO, to assist services in autonomously adjusting their own resource allocations.
{\sysname} conveys this bridging ``assistance'' through
\textit{performance targets}, which translate the desired application performance to local
proxy metrics measurable by services. Doing so hides low-level resource control
details from the SLO feedback mechanism. In this paper, we use CPUs to discuss
the framework design---not only is the CPU harder to manage due to its higher usage
fluctuation over time~\cite{alibaba_microservices, harvest_vm, deathstarbench},
but it also has an immense impact on microservice response
time~\cite{microservices_factors, sinan, smartharvest}.

At each microservice, {\sysname} locally runs a lightweight resource controller
called \textit{{\syslocal}}. {\syslocal} swiftly adjusts CPU allocations through
OS APIs (e.g., CPU quota in Linux's cgroups), to ensure its governed
microservice reaches the given performance target. {\sysname} represents this
target using an unconventional metric---\textit{CPU throttles}, namely the
number of times a service exhausts its CPU quota in a time period. Not only are
CPU throttles sufficiently cheap to sample at high frequency to enable
{\syslocal}s' timely adjustments, but we also observe that they have higher
correlation with latencies than other proxy metrics such as CPU utilization
(\S\ref{subsec:microbenchmarks}).
These characteristics make CPU throttles an indicative target to track locally,
for maintaining an end-to-end SLO.
At the application level, \sysname employs a centralized SLO
feedback controller called \textit{{\sysglobal}}.
It observes the application workload measured by RPS
(requests per second) and learns to determine the most cost-effective
performance targets that maintain the SLO,
using a lightweight online algorithm
known as contextual bandits~\cite{cb_bakeoff}.

This paper makes the following key contributions:
\begin{itemize}[itemsep=0pt,topsep=3pt,leftmargin=*]
  \item We examine unique implications that SLO-targeted microservices introduce
  to resource management (\S\ref{sec:motivation}). Since there is no strong
  correlation between the end-to-end application performance and per-service
  resource usage, directly computing the optimal resource
  allocations is non-trivial.

  \item {\sysname} is a bi-level learning-assisted framework
  (\S\ref{sec:design}), to embrace distinct levels of distributed system
  behavior. It separately designs mechanisms of application-level SLO feedback
  and service-level resource control, and introduces CPU-throttle-based
  performance targets to bridge them.

  \item Comprehensive experiments (\S\ref{sec:eval})
  demonstrate {\sysname}'s superior CPU savings over state-of-the-art
  heuristics and ML-based baselines, in three SLO-targeted applications:
  Train-Ticket~\cite{trainticket}, Social-Network~\cite{sinan}, and
  Hotel-Reservation~\cite{deathstarbench}. Compared with the best-performing
  baseline in each application, {\sysname} maintains the SLO for the
  99th percentile latency while
  saving up to 26.21\% CPU cores for Train-Ticket, up to 25.93\% for
  Social-Network, and up to 7.34\% for Hotel-Reservation, across four
  real-world workload patterns. Finally, running
  Social-Network over a 21-day period with production workloads from a global
  cloud provider, {\sysname} saves up to 35.2 CPU cores while reducing
  hourly SLO violations by 13.2$\times$.
\end{itemize}

\tightsection{Background and Motivation}
\label{sec:motivation}

Our goal of SLO-targeted resource
management is to minimize the total CPU allocations to microservice-based
applications, while avoiding SLO violations on the user request latency.
Following real-world findings~\cite{realworld_slo}, our SLO is an upper
limit on tail latencies, specifically the 99th percentile (P99) request
latencies unless otherwise noted.

\subsection{Implications of microservices}
\label{sec:implications}

Unlike monolithic applications, the distributed nature of microservices implies
that multiple services collectively contribute to the end-to-end latency. This
section presents observations, to motivate its implications on computing
per-service resource allocations from the end-to-end latency SLO.

\subsubsection{Service execution dependencies}
\label{sec:cascading_effects}

As user requests traverse services, their end-to-end latency is a function of
per-service performance (and hence resource usage). Being
functionally different, services can consume resources differently.
Moreover, service execution dependencies can introduce complex correlations
to this function---not only are there various patterns
such as parallelism, but services can also exhibit unexpected increases in
resource demand.

An illustrative example is backpressure~\cite{deathstarbench}---as an
under-provisioned service undergoes performance degradation during request
processing, the resource manager can misinterpret its idling parent's
longer response time as the culprit. Simply
identifying all parent-child relations does not fully solve the problem,
as backpressure can vary subtly depending on service implementations.
In one case we encountered, the CPU usage of a waiting
parent unexpectedly increased with the number of requests, which was
counterintuitive as waiting for child services should result in
idle CPUs. Further investigations revealed that the parent service
spawned a separate thread for each outstanding request
(i.e., Thrift's \texttt{TThreadedServer} RPC model),
leading to excessive
thread maintenance and spurious context switching.
An alternative implementation with non-blocking or asynchronous I/O
(e.g., Thrift's \texttt{TNonBlockingServer}) eliminated the problem.

To grapple with the complexities arising from service dependencies,
prior work considers how the end-to-end performance
\textit{directly} correlates with application-wide
execution dependencies. However, maintaining an accurate and up-to-date
global view of these dependencies is both challenging and costly.
First, the interdependencies among services are constantly evolving
during development, often with multiple versions of the same service
coexisting~\cite{seshagiri2022sok}, requiring frequent updates to the global view.
Second, although ML-based resource management strategies~\cite{sinan,graf} have the potential
to comprehend complex and large-scale service
dependencies,
they may entail substantial training and retraining expenses.
Third, fine-grained distributed tracing (beyond
the basic monitoring through sampling-based tracing) may be necessary for
resource managers to observe and analyze service dependencies~\cite{firm, seer},
resulting in additional system overheads from increased instrumentation
and telemetry collection.

\begin{tcolorbox}[
  width=\linewidth,
  top=0pt, bottom=-8pt,
  left=8pt, right=8pt,
  colback=white, colframe=white]
  \textit{\textbf{Observation \#1:} Maintaining an up-to-date global view of service
  dependencies can be impractical.}
\end{tcolorbox}

\subsubsection{Delayed end-to-end performance feedback}
\label{sec:delayed_effects}

The chain of service execution dependencies brings about the \textit{delayed
effect}, i.e., a time delay for the impact of any changes in resource
allocations or workloads to be fully observed in the end-to-end performance.
This prevents resource managers from immediately responding to
misallocations.
The delayed effect is often amplified. One source is service queues---under-provisioning of resources will cause requests to
accumulate in queues, and thus SLO violations are not detected until all
queued requests are eventually processed or timed out.
Even if resources are scaled at this point, it takes time to flush
queues~\cite{deathstarbench,sinan}. Another amplification is that
SLO is typically defined on
aggregated performance data (e.g.,
percentiles), which require a sufficient number of requests to be profiled.

In light of the delayed effect, prior work~\cite{sinan} proposes to
proactively predict the long-term impact of resource changes on the
end-to-end performance.
Such performance predictions are theoretically possible
but they usually involve expensive data collection and model
training. On the other hand, prematurely deploying ML models can result in
a high percentage of mispredictions.
For instance, our efforts to fully train Sinan's neural
networks~\cite{sinan} for a
28-microservice application took 14+ hours, plus
$\sim$6 hours to collect 20,000 training data points.
Despite reproducing the published prediction accuracy, we observed that
mispredictions can trick resource managers to overallocate at
least 40.75\% more CPU cores (\S\ref{sec:eval}).

\begin{tcolorbox}[
  width=\linewidth,
  top=0pt, bottom=-8pt,
  left=8pt, right=8pt,
  colback=white, colframe=white]
  \textit{\textbf{Observation \#2:} Predicting end-to-end application performance under
  the delayed effect can be unreliable.}
\end{tcolorbox}

\subsection{A practical approach}
\label{sec:principles}

In light of the observations in \S\ref{sec:implications}, a more promising
approach for SLO-targeted resource managers is to embrace the distributed
nature of microservices by taking into account of the distinct levels of
system behavior---the macro perspective reveals the end-to-end
performance (e.g., user request latencies) and SLO,
whereas the micro perspective is scoped to local measurements (e.g., service CPU usage) and control.

Naturally, these two levels can map to:
\textit{\textbf{(1)}} \textit{application-level SLO feedback}, which
compares the end-to-end performance and SLOs, and \textit{\textbf{(2)}}
\textit{service-level resource control}, which computes resource allocations
based on local measurements. In fact, if we architecturally decouple these
mechanisms, it becomes feasible to position them close to their required inputs.
Doing so brings the benefit of fast reaction, which opens up opportunities for
resource managers to relax the requirement of computing the optimal resource
allocations. Rather than striving to accurately model service dependencies
(\S\ref{sec:cascading_effects}) or predict long-term application-wide behavior
(\S\ref{sec:delayed_effects}), we can now employ lightweight service-level
controllers that autonomously and swiftly adjust resource allocations,
assisted by periodic guidance computed at the application level
through a lightweight online learning approach.

In summary, SLO-targeted resource managers for microservices should incorporate
the following design principles.

\begin{enumerate}[itemsep=0pt,topsep=3pt,leftmargin=*]
  \item Decouple mechanisms of application-level SLO feedback and service-level resource control.
  \item Rapidly drive per-service resource control with local performance targets and near-term prospects.
  \item Achieve practicality through lightweight solutions.
\end{enumerate}

\tightsection{The {\sysname} Framework}
\label{sec:design}

Following the design principles laid out in \S\ref{sec:principles}, we present {\sysname}, a practical and readily deployable
resource management framework for SLO-targeted microservices.

\subsection{Overview}
\label{subsec:design_overview}

\sysname is a bi-level learning-assisted framework,
consisting of an application-wide global controller and
per-service local controllers.
The application-level controller is based on online learning, periodically
assisting local resource control with its visibility into application
workloads, end-to-end latencies, and SLO violations.
The service-level controllers, on the other hand, are heuristic-based,
continuously performing fast and fine-grained CPU scaling using local metrics as well as the assistance from the global controller.

The ``assistance'' bridging the two levels is based on the notion of
\textit{performance target}, a target performance level set by the
application-wide controller for per-service controllers to
attain.
{\sysname} implements the performance target with
\textit{CPU throttle ratio}---the fraction of time
a microservice is stopped by the underlying CPU scheduler.
This design is motivated by the strong correlation between CPU throttles
and service latencies revealed by our correlation test
(\S\ref{subsec:microbenchmarks}).
Maintaining a CPU throttle ratio locally also allows
tolerating a certain range of workload fluctuations
(\S\ref{subsec:microbenchmarks}).
When per-service controllers fail to rein in
end-to-end latencies (e.g., the workload exceeds the tolerable range),
the application-wide controller issues lower throttle targets
to guide local controllers to allocate more CPUs.
Conversely, higher throttle targets are assigned in the event of
CPU over-provisioning.

Acting as a broker, the performance target allows
{\sysname} to decouple the mechanisms of application-level SLO feedback and
service-level resource control. Consequently, we are able to simplify the learning
process of the application-level controller by concealing low-level resource details
and avoiding the overhead of aggregating them,
while enabling per-service controllers to focus on a self-contained in-situ
task---reaching a given performance target using locally available information.
Our bi-level design sets us apart from approaches
that directly infer resource demands with proxy metrics
(e.g., \cite{k8s_autoscale}) or machine learning (e.g., \cite{sinan}).

\begin{figure}[t]
    \centering
    \includegraphics[width=\columnwidth]{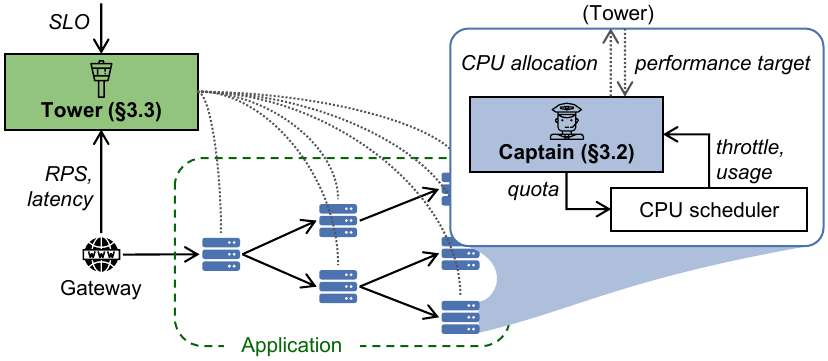}
    \caption{{\sysname} features bi-level resource
    management: The application-level learning-based controller (\sysglobal),
    observing end-to-end latencies and workloads, periodically
    sets performance targets, expressed as CPU throttle ratios,
    for per-service heuristic controllers (\syslocals) to meet.
    }
    \label{fig:arch}
\end{figure}

Figure~\ref{fig:arch} depicts the architecture of {\sysname}.
We refer to the per-service controllers as \textit{{\syslocals}}
(\S\ref{subsec:local_plane}), and the application-wide controller
as the \textit{{\sysglobal}} (\S\ref{subsec:global_plane}).\footnote{
The air traffic control
``tower'' (application-wide controller) assigns ``routes''
(performance targets) to flights, while each ``captain''
(per-service controller) follows the assigned route by actually steering
the aircraft.}

\para{{\sysname} {\syslocals}.} At the local level, each microservice runs a
{\syslocal} instance, which periodically receives performance targets---CPU throttle ratios---from the
{\sysglobal} and strives to realize these targets using heuristic control.
The heuristic control algorithm collects statistics on CPU usage and throttles,
and employs two feedback control loops
to scale CPUs up and down, respectively.
This lightweight design ensures swift and fine-grained CPU autoscaling of the
\syslocal even amid rapidly fluctuating workloads.

\para{{\sysname} {\sysglobal}.} At the global level, the {\sysglobal}
leverages \textit{contextual bandits}~\cite{cb_bakeoff},
a lightweight class of online reinforcement learning (RL),
to dynamically determine suitable performance targets that
maintain the SLO. It monitors application workload (e.g., RPS)
and observes CPU allocations and end-to-end latencies (along with
associated SLO violations) as feedback for its output targets.
This online learning approach is directly applicable to any microservices,
eliminating the need for extensive offline profiling or training.

Overall, {\sysname} takes a pragmatic stance and provides
a resource management framework that is readily deployable across
diverse latency-sensitive  microservice applications.
Next, we elaborate on {\sysname} from
the bottom up, starting off with {\syslocals} (\S\ref{subsec:local_plane}),
followed by the {\sysglobal} (\S\ref{subsec:global_plane}).

\subsection{Per-service controllers---\textit{{\syslocals}}}
\label{subsec:local_plane}

Each {\syslocal} periodically (e.g., every minute)
receives a target CPU throttle ratio from the {\sysglobal}.
Given a throttle target, {\syslocal}
focuses on a self-contained, in-situ task---scaling up and down
the CPUs made available to its governed service
in order to meet the throttle target upon changing demand.
Algorithms~\ref{alg:local-scale} and~\ref{alg:local-rollback} present the
pseudocode of {\syslocal}'s main components, which we describe in detail
below.

\subsubsection{Resource metrics and knobs}
\label{subsec:resource_metrics_knobs}

Common CPU schedulers in the OS,
such as the Linux CFS scheduler we use as a running example in this paper,
assign each microservice
a CPU quota (e.g., \texttt{cpu.cfs\_quota\_us})
to limit and isolate resource usage.
To accomplish the task of maintaining a target CPU throttle ratio,
{\syslocal} continuously collects two statistics exposed by the OS
in each time window---CPU throttle count and CPU usage.

\para{CPU throttle count.} The Linux CFS scheduler
maintains a CPU throttle count for each microservice in the variable
\texttt{cpu.stat.nr\_throttled}, which represents the cumulative number
of CFS periods (100 ms by default) during which the CPU quota has been exhausted.
Intuitively, if the CPU quota is used up early in a CFS period
before a request can be fulfilled, the request will be
approximately delayed by the remaining period,
underscoring the importance of avoiding CPU throttles
when maintaining latency SLOs.
Anecdotal evidence in blog posts~\cite{indeedblog,ibmblog} corroborates our intuition.
To calculate the CPU throttle ratio over a time window,
we divide the increase in the CPU throttle count
(\texttt{cpu.stat.nr\_throttled}) by the number of elapsed CFS periods.

\para{CPU usage.} The Linux CFS scheduler also reports
the total CPU time consumed by
a microservice as \texttt{cpuacct.usage}. This metric
is particularly useful when the CPU is \textit{over-provisioned}, as it reveals
the actual (lower) CPU demand. Otherwise, this actual demand would be capped
by the allocated CPUs if under-provisioned and thus remain unknown.

\setlength{\floatsep}{5pt}
\setlength{\textfloatsep}{15pt}
\SetInd{0.1em}{0.45em}
\begin{algorithm}[t]
    \small
    \caption{{\syslocal}: scaling up and down}
    \label{alg:local-scale}
    \SetKwData{N}{N}
    \SetKwData{Quota}{quota}
    \SetKwData{NewQuota}{proposed}
    \SetKwData{Margin}{margin}
    \SetKwData{ThrottleCount}{throttleCount}
    \SetKwData{ThrottleRatio}{throttleRatio}
    \SetKwData{UsageHistory}{history}
    \SetKwData{Target}{throttleTarget}
    \SetKwData{ScaleDownStepMin}{scaleDownStepMin}
    \SetKwData{ScaleDownStepMax}{scaleDownStepMax}
    \SetKwFunction{Max}{max}
    \SetKwFunction{Std}{stdev}
    \CommentSty{/* executes every $N$ periods */}\\
    $\ThrottleCount =$ throttle count during last $N$ periods\;
    $\ThrottleRatio = \ThrottleCount / N$\;
    $\Margin = \Max{$0$, $\Margin + \ThrottleRatio - \Target$}$\;
    \eIf{$\ThrottleRatio > \alpha \times \Target$}{
        \label{alg:local-scale:scale-up-begin}
        \CommentSty{/* multiplicatively scale up */}\\
        $\Quota = \Quota \times (1 + \ThrottleRatio - \alpha \times \Target)$\;
        \label{alg:local-scale:scale-up-end}
    }{
        \CommentSty{/* instantaneously scale down */}\\
        \label{alg:local-scale:scale-down-begin}
        $\UsageHistory =$ CPU usage history in the last $M$ periods\;
        $\NewQuota = \Max{\UsageHistory} + \Margin \times \Std{\UsageHistory}$\;
        \If{$\NewQuota \leq \beta_{\max} \times \Quota$}{
            $\Quota = \Max{$\beta_{\min} \times \Quota,\ \NewQuota$}$\;
        }
        \label{alg:local-scale:scale-down-end}
    }
\end{algorithm}
\begin{algorithm}[t]
    \small
    \caption{{\syslocal}: rollback mechanism}
    \label{alg:local-rollback}
    \SetKwData{N}{N}
    \SetKwData{Quota}{quota}
    \SetKwData{LastQuota}{lastQuota}
    \SetKwData{Margin}{margin}
    \SetKwData{ThrottleCount}{throttleCount}
    \SetKwData{ThrottleRatio}{throttleRatio}
    \SetKwData{Target}{throttleTarget}
    \CommentSty{/* executes every period for $N$ periods after each scale-down */}\\
    $\LastQuota =$ CPU quota before scale-down\;
    $\ThrottleCount =$ throttle count since scale-down\;
    $\ThrottleRatio = \ThrottleCount / N$\;
    \If{$\ThrottleRatio > \alpha \times \Target$}{
        \CommentSty{/* revert to the previous (higher) quota before scale-down}\\
        \CommentSty{\quad with an additional allocation equal to the quota difference */}\\
        $\Quota = \LastQuota + (\LastQuota - \Quota)$\;
        $\Margin = \Margin + \ThrottleRatio - \Target$\;
    }
\end{algorithm}

\subsubsection{Multiplicative scale-up}
\label{subsubsec:scale_up}

In every time window of $N$ ($N=10$ by default) CFS periods,
each \syslocal compares the measured CPU throttle ratio at its microservice
with the target ratio.
When the measured ratio exceeds the target, it indicates the CPU is
under-provisioned, demanding a prompt increase in the CPU quota to prevent
imminent SLO violations at the application level.

To ensure that any desired target can be reached quickly within several steps,
{\syslocal} increases the current CPU quota \textit{multiplicatively}.
We further make the size of the increase proportional to
the difference between the measured CPU throttle ratio and the target ratio.
This represents a form of proportional control, where a larger difference
results in a larger stride in the CPU quota increase.
The rationale is that when the difference is significant, a queue of requests is
likely to have built up, thus requiring more CPUs to drain.

In practice, we find that the local workload arriving at a microservice
is naturally bursty and irregular---regardless of the pattern of end-to-end
requests---tricking {\syslocals} into spurious scale-ups.
Hence, we execute the scale-up only when the CPU throttle ratio
surpasses ``$\alpha \times \text{target ratio}$'' ($\alpha \ge 1$),
where $\alpha$ is a customizable weight that controls the sensitivity to
transient load spikes. Correspondingly, the CPU quota is also multiplied by
``$1 + \text{throttle ratio} - \alpha \times \text{target ratio}$'' in each step.
The pseudocode is in Line~\ref{alg:local-scale:scale-up-begin}--\ref{alg:local-scale:scale-up-end}
of Algorithm~\ref{alg:local-scale}.

\subsubsection{Instantaneous scale-down}
\label{subsubsec:scale_down}

Under frequent CPU throttling, {\syslocal} is forced to incrementally probe the
actual CPU demand of the service.
In contrast, when the measured throttle ratio is below the target ratio,
the service's CPU demand has been adequately met.
Consequently, historical CPU usage begins to more accurately reflect
the actual (less throttled) CPU demand and help \textit{instantaneously}
determine the desired CPU quota.

Motivated by this characteristic of over-provisioning,
the {\syslocal} maintains a sliding window of CPU usage over the most recent
$M$ ($M=50$ by default) CFS periods, and calculates a new CPU quota based on
two statistics from the sliding window: the maximum and the standard deviation of CPU usage.
Specifically, the proposed quota is ``max CPU usage $+$ $margin \times \text{standard deviation of CPU usage}$,'' where $margin \ge 0$ is a
dynamically tuned parameter that generally increases when the CPU throttle
ratio exceeds the target ratio and decreases otherwise.
To avoid unnecessary fluctuations in CPU allocation,
the proposed quota is put into action only when it
represents a significant-yet-moderate change.
The details are described in
Line~\ref{alg:local-scale:scale-down-begin}--\ref{alg:local-scale:scale-down-end}
of Algorithm~\ref{alg:local-scale}.

Our scale-down design draws inspiration from prior work~\cite{autopilot,rubas},
but differs
in the carefully maintained parameter $margin$ that depends on CPU
throttles. Intuitively, if the CPU
is recently throttled more often than desired,
we should be more conservative by using a larger $margin$ in the subsequent
scale-down to avoid overreacting to momentary tranquility amid workload
spikes; and vice versa.
In summary,
historical CPU usage in the sliding window allows for instantaneous
scale-down, reclaiming extra CPU allocations in a single step.

\subsubsection{Rollback mechanism after scaling down}
\label{subsubsec:rollback}

Accidentally scaling up CPUs only leads to resource waste (and existing
cloud applications tend to be over-provisioned); however, mistakenly
scaling down the CPU allocation to any microservice may cause SLO violations at
the application level. Thus, we introduce a fast rollback mechanism to the
{\syslocal} to revert ``reckless'' scale-downs as follows.

After each scale-down, we continuously check whether it is
``reckless''---if it has caused the CPU throttle ratio to
exceed $\alpha \times \text{target ratio}$---during
\textit{every} CFS period within the next $N$ periods.
We note that the triggering condition is the same as that used for scaling up,
but due to the urgency of initiating a rollback, this check is performed more
frequently, without waiting for the
\syslocal's regular decision-making interval ($N$ periods).
After a rollback is triggered, the current CPU quota is restored to the
previous (higher) quota used before the scale-down, plus an additional allocation
equal to the difference between the two quotas.
We grant slightly more CPUs to account for the potential processing delays that may have
occurred since the erroneous scale-down.
Details of the rollback mechanism are presented in Algorithm~\ref{alg:local-rollback}.

\subsection{Application-level controller---\textit{{\sysglobal}}}
\label{subsec:global_plane}

In {\sysname}, {\sysglobal} delegates the in-situ resource control to
per-service {\syslocals} and only provides periodic assistance by dispatching the target
CPU throttle ratios for \syslocals to meet. This indirection via throttle targets
effectively avoids the latency overhead associated with distributed tracing and
logging, while retaining {\sysglobal}'s global perspective on end-to-end
requests and SLO feedback.

We design {\sysglobal} to compute a new target throttle ratio
infrequently, e.g., once a minute, leaving ample time for tail request latencies
and average CPU usage to stabilize as the new target settles in.
Importantly, doing so minimizes the influence of \sysglobal's previous decisions,
simplifying the problem into a ``one-step'' decision-making process:
\sysglobal only needs to determine the optimal CPU throttle targets
for the current step, without considering their long-term consequences.

This ``one-step'' nature motivates us to employ \textit{contextual bandits},
a lightweight class of online reinforcement learning (RL) algorithms.
In the taxonomy of RL, contextual bandits can be viewed as
one-step RL and are well suited for real-time online scenarios
in which the algorithm is required to learn efficiently from a limited amount of
sample data.

\subsubsection{Primer on contextual bandits}
\label{subsubsec:contextual_bandits}

Recent work has modeled resource management with sequential decision-making
paradigms and seen the application of multi-armed
bandits~\cite{cola,autopilot} and reinforcement
learning~\cite{firm,decima,genet}.  Contextual bandits are intermediate between
multi-armed bandits and the full-fledged RL~\cite{sutton2020rl}.

Contextual bandits are like multi-armed bandits in that they are well suited
to problems where an \textit{action} (e.g., CPU throttle targets) taken
at a step (e.g., one-minute interval) does not have long-term
impact beyond that step. They receive a \textit{cost}
(negative reward) as feedback for the chosen action, and aim to minimize
the cumulative cost (e.g., comprising CPU allocations and SLO
violations).
Conversely, contextual bandits also differ from multi-armed bandits by
their ability to make decisions based on the observation of the system state,
known as the \textit{context} (e.g., RPS). This context can provide valuable
information that aids in the learning process
(\S\ref{subsec:microbenchmarks}).

In contrast to the full RL, which optimizes a sequence of future steps,
contextual bandits only optimize the current step owing to their
assumption that each chosen action only affects the immediate outcome
without long-term consequences. Moreover, full RL typically demands
extensive offline training before deployment as well as frequent retraining
(e.g., upon significant changes in microservices), whereas contextual bandits
are more lightweight (with simpler models) and suitable for online learning with considerably fewer
samples.

In solving contextual bandit problems, a common approach is to train a
cost-prediction model that estimates the cost of taking each action
within a context. Due to their inherent partial observability, however, contextual
bandits can only observe the costs of actions they select but not the costs of
others.
To enhance their performance and sample efficiency, a widely adopted improvement is
to estimate the costs of unused actions via counterfactual
estimates~\cite{dudik2011doubly,schnabel2016recommendations,agarwal2014taming}.
This approach reduces contextual bandit problems to cost-sensitive
classification~\cite{langford2007epoch}, which can then be addressed using
standard supervised learning. We adopt this approach and refer the reader to
Bietti et al.~\cite{cb_bakeoff} for more details.

\subsubsection{Realizing contextual bandits in {\sysglobal}}
\label{subsubsec:tower_cb}

Next, we describe the contextual bandit algorithm used in {\sysglobal}.
The algorithm operates with a step size of one minute, and it aims to learn
to output an action that incurs the lowest cost given the observed context
at each step.

\para{Context.} {\sysglobal} selects the average RPS observed in the last
step as the context because the optimal CPU throttle
target depends on the RPS (\S\ref{subsec:microbenchmarks}).
We refrain from predicting the RPS for the next step
due to the inherent difficulty in accurately forecasting RPS; moreover,
our {\syslocals} have been intentionally designed to tolerate short-term RPS
fluctuations (\S\ref{subsec:microbenchmarks}).
Other metrics such as CPU usage are not included in the context as they are
merely the byproducts of applying a throttle target to an RPS,
with the RPS serving as the primary causal factor.
The composition of the workload (i.e., the distribution of different request types) is relevant,
but our focus in this work remains on
constant workload composition (Appendix~\ref{sec:appendix_workload}), following the setup in prior work~\cite{deathstarbench,sinan}.

\para{Action.} Given an instantiation of the {\syslocal}'s resource control algorithm,
we search for a ladder of CPU throttle targets as the actions.
The search is a one-time process for all applications.
By default, our action space consists of 9 throttle targets, ranging from 0 to 0.3 (\S\ref{sec:impl}).

\para{Reduction of action space.} A microservice-based application can contain
10--1000s of services~\cite{twitter_microservices, netflix_microservices,
uber_microservices, alibaba_microservices, wechat_overload}. In the case of 9 throttle targets,
generating a different CPU throttle target for each individual service would
result in $9^{\#\text{services}}$ actions, rendering it infeasible
for contextual bandits to learn.
As a solution, {\sysglobal} clusters microservices into two classes and outputs
an action for each class, effectively reducing the action space to $9^2=81$.
To implement the clustering, we use the standard $k$-means
algorithm~\cite{k_means} to group microservices based on their average CPU usage.
Our empirical results in \S\ref{subsec:microbenchmarks} suggest a diminishing
return beyond two clusters.

\para{Cost function.} We define the cost received per step as follows.
When the SLO is met after the step, we only use the total CPU
allocation as the cost, since the actual
latencies below SLO matter no more.
To this end, {\sysglobal} requests {\syslocals} to send their actual CPU allocations
as feedback every minute, and then normalizes the total allocation
linearly into $[0,1]$.
On the other hand, when the SLO is violated,
we set the cost to only contain the tail latency,
linearly normalized to $[2,3]$ considering the higher priority of
SLO violations.
We arrived at the two normalization ranges above based on their empirical
performance compared with other ranges we tested, but we do not claim our cost
function is the best.

\para{Noise reduction for costs.} Our contextual bandit algorithm learns online
and updates its model weights on every (context, action, cost) tuple,
i.e., most recent RPS, two throttle targets, and the incurred cost.
In reality, however, we observe highly noisy costs that result in confusion and poor
performance of the model, supposedly due to the
complex microservice system and the dynamics in {\syslocals}.
To address this, we buffer and group recent samples using (context, action)
as the index after quantizing the RPS.
Given a new sample, we use the median cost of the group it falls into---rather than
the cost computed for that individual sample---to update the
model. Doing so significantly reduces the noise in costs and
stabilizes the online learning process.

\para{Exploration.} Similar to multi-armed bandits and RL, contextual
bandits rely on exploration to acquire knowledge about
the costs of different actions, e.g., using
$\epsilon$-greedy~\cite{langford2007epoch} to choose a
random action with a small probability of $\epsilon$
(the best action is selected otherwise).
Despite a reduced action space, randomly exploring all 81 actions within a context
remains inefficient as each sample requires one minute to complete;
repeated sampling is further required to calibrate noisy cost estimates.
To ensure efficient exploration without impeding online learning,
we explore only the neighbors of the best action in the action space.
Given a sorted ladder of the available CPU throttle target ,
$r_1<r_2<\ldots<r_9$,
if the best action consists of ($r_i$, $r_j$),
$1 \le i,j \le 9$, then each of its neighbors ($r_i$,
$r_{j-1}$), ($r_i$, $r_{j+1}$), ($r_{i-1}$, $r_j$), ($r_{i+1}$, $r_j$) is
explored next with an equal probability of $\epsilon / 4$ (subject to boundary conditions).
The rationale is that the throttle target ladder is monotonic, allowing
{\sysglobal} to move upward or downward one step at a time without missing the
optimal action.

\tightsection{Implementation}
\label{sec:impl}

Our current implementation supports microservice applications
deployed as pods on Kubernetes, but it can be easily extended to other container
orchestration frameworks (e.g., OpenShift and Docker Swarm). {\sysname} is open-sourced at \url{https://github.com/microsoft/autothrottle}.

\vspace{0.5\baselineskip}

\para{{\syslocal}.}
Each microservice is associated with a {\syslocal} co-located on the same
worker node, so we deploy {\syslocals} as processes on worker
nodes of the Kubernetes cluster. {\syslocal} implements the following three
functionalities. First, it communicates with the {\sysglobal}
over a TCP socket, exchanging CPU throttle targets and allocations.
Second, it collects CPU throttling and usage
statistics from Linux cgroup API in every CFS period of 100 ms, as the input
to the local resource controller.
Third, it runs the resource controller for all microservices on the
same worker node, and sets their CPU quotas (\texttt{cpu.cfs\_quota\_us}) accordingly.
As {\syslocal} only comprises lightweight heuristic-based
control loops, it does not require any pre-deployment training.

The pseudocode of \syslocal is outlined in
Algorithms~\ref{alg:local-scale} and \ref{alg:local-rollback}. Our default
parameters are $N = 10$, $M = 50$, $\alpha = 3$, $\beta_{\max}
= 0.9$, $\beta_{\min} = 0.5$. They can be adjusted accordingly. A larger $N$ or $M$
lowers sensitivity to the noise in CPU usage, hence slower reaction.
$\alpha$ sets the supported range of throttle ratios to (0, $1/\alpha$).
A smaller $\alpha$ increases the upper bound but decreases the tolerance on
throttle ratio fluctuations.
$\beta_{\max}$ and $\beta_{\min}$ prevent overly small or large allocation changes.

\para{{\sysglobal}.}
One instance of {\sysglobal} runs globally alongside the application (i.e., in
the same cluster), initialized with a user-specified SLO.
It collects average RPS and tail latencies from
the Locust workload generator, but can be extended to hook up to an
application gateway.
Furthermore, {\sysglobal} receives the actual CPU allocations from {\syslocals}
after dispatching CPU throttle targets to them every minute.

{\sysglobal} leverages the widely used Vowpal Wabbit (VW) library~\cite{vw}
to implement contextual bandits.
For each group of microservices,
the model outputs one of the 9
throttle targets: 0.00, 0.02, 0.04, 0.06, 0.10, 0.15, 0.20,
0.25, and 0.30.
Designed for efficient online learning,
VW offers lightweight model options such as
linear regression or a shallow neural network with a single hidden layer.
We opt for a neural network model with 3 hidden units
after performing an ablation study (\S\ref{subsec:microbenchmarks}),
and train it with a learning rate of 0.5.
The doubly robust estimator~\cite{dudik2011doubly} is employed
in the bandits for policy evaluation to estimate the costs of
untaken actions.
Moreover, we disable the native $\epsilon$-greedy algorithm
to implement our customized exploration strategy (\S\ref{subsubsec:tower_cb}).
The specific VW usage is detailed in Appendix~\ref{sec:appendix_vw}.

Online training starts with an exploration stage, which allows VW to randomly
explore how different CPU throttle targets would impact application
latencies. During this stage, each randomly chosen action will be executed for
2 minutes. Only the second minute is used for cost calculation and training,
in order to avoid interference from the previous chosen action.
This exploration stage lasts $\sim$6 hours, during which application
latencies may exceed the SLO.

After the exploration stage, {\sysglobal} starts to exploit the best action,
while still exploring neighboring actions with a total of 10\% probability using
$\epsilon$-greedy. {\sysglobal} runs every minute to collect last minute's
(context, action, cost) sample. All recent samples are grouped using
(context, action) as the index with RPS quantized into bins of 20,
and each group's cost is defined as the median cost of the group.
Since training each unique (context, action) only once is
insufficient for contextual bandits, 10,000 training data points are
sampled from these groups randomly. A contextual bandit model is then
trained on these samples, and predicts the next best action based on RPS.
As a reference, this training-and-prediction process takes less than one second in our
setup.

\tightsection{Evaluation}
\label{sec:eval}

We evaluate {\sysname}'s superior resource saving with three SLO-targeted
microservice applications, against state-of-the-art heuristic- and ML-based
baselines. Major results include:

\textbf{\textit{(1)}} Over the best-performing baseline in each application,
{\sysname} maintains the given application P99 latency SLO, while achieving a
CPU core saving up to 26.21\% for Train-Ticket, up to 25.93\% for Social-Network, and
up to 7.34\% for Hotel-Reservation. Over all baselines, its savings can be up to
93.84\%, 55.32\%, and 83.99\%, respectively.

\textbf{\textit{(2)}} A 21-day study of Social-Network (with real-world workload
trace from a global cloud provider) shows a saving up to 35.2 CPU cores, over
the best-performing baseline. Meanwhile, it reduces hourly SLO violations from
71 to 5.

\textbf{\textit{(3)}} Microbenchmarks evaluate {\sysname}'s design and tolerance
to workload fluctuations and load-stressing.

\subsection{Methodology}
\label{sec:eval_methodology}

\para{Benchmark applications.}
We deploy three SLO-targeted microservice applications: \textit{\textbf{(1)}}
Train-Ticket~\cite{trainticket}, with 68 distinct services,
\textit{\textbf{(2)}} Hotel-Reservation from DeathStarBench~\cite{deathstarbench},
with 17 distinct services, and \textit{\textbf{(3)}}
Social-Network used in Sinan~\cite{sinan}, a variant of the
Social-Network application from DeathStarBench, with
28 distinct services including two ML inference serving services:
a CNN-based image classifier and an SVM-based text classifier.
These applications are representative of real-world
microservices, with stateless services
(e.g., business logic), data services (e.g., key-value stores), and gateways.
Deployments are managed by Docker and
Kubernetes. Parent-child service communications are through popular RPC
frameworks such as gRPC and Thrift.

Application SLOs are specified on the hourly P99
latency~\cite{realworld_slo}---1,000~ms for Train-Ticket, 200~ms for Social-Network,
and 100~ms for Hotel-Reservation.

\vspace{0.5\baselineskip}
\para{Comparison baselines.}
Baselines include \textit{\textbf{(1)}} Kubernetes default
autoscalers~\cite{k8s_autoscale}
(denoted as ``K8s-CPU'' and ``K8s-CPU-Fast''), and
\textit{\textbf{(2)}} state-of-the-art ML-driven solution, Sinan~\cite{sinan}.

K8s-CPU locally maintains each service's average CPU utilization, with respect
to the user-specified CPU utilization threshold (e.g., 50\%). Every $m$$=$15
seconds, it measures service's CPU usage, and computes the optimal allocation by
``CPU usage / CPU utilization threshold.'' Then, it sets the CPU limit to
the largest allocation computed in the last $s$$=$300 seconds. We also include a
faster version called K8s-CPU-Fast, which has $m$$=$1 and $s$$=$20.
Since Kubernetes relies on users to properly translate the application SLO to
CPU utilization threshold, we manually try different thresholds to find the
appropriate one for each experiment (Appendix~\ref{sec:appendix_k8s_cpu_targets}).

Sinan leverages ML models (e.g., a convolutional neural network and
a boosted tree model)
to globally assess each service's resource allocation. Starting with the
open-sourced Sinan~\cite{sinan_repository}, we follow instructions to train
application-specific models offline for 20+ hours. Since Sinan relies on
users to properly set several hyperparameters, we manually tune for each
application. During experiments, we run Sinan every second---given
historical resource usage and latencies, Sinan tries to predict the optimal CPU
allocation that is unlikely to violate the SLO over both the short and long
terms.

\begin{figure}[t]
  \begin{subfigure}{0.49\columnwidth}
    \includegraphics[height=66pt]{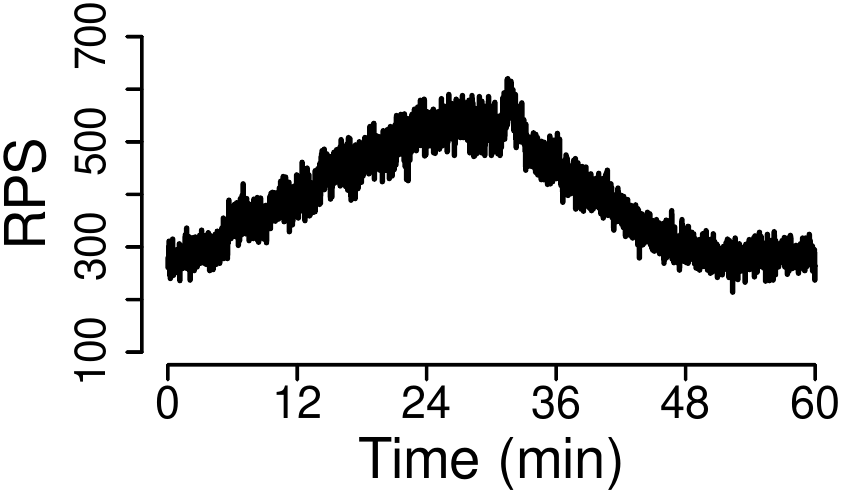}
    \caption{Diurnal}
    \label{fig:traces_diurnal}
  \end{subfigure}
  \hfill
  \begin{subfigure}{0.49\columnwidth}
    \includegraphics[height=66pt]{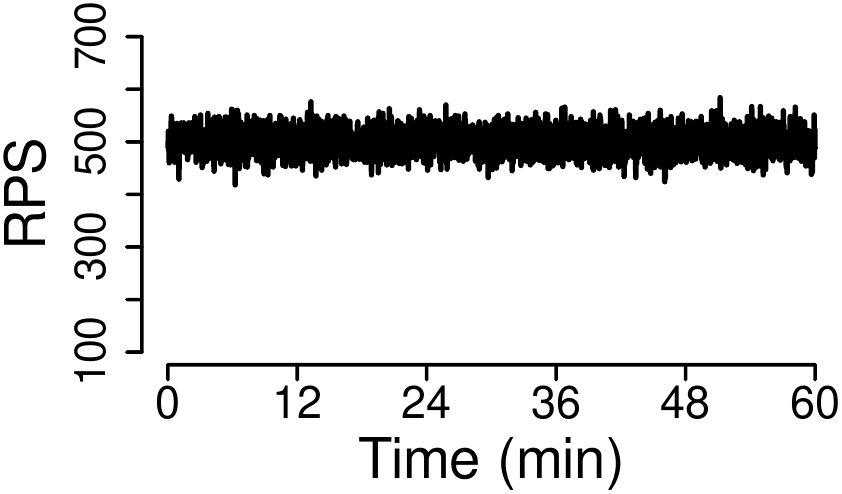}
    \caption{Constant}
    \label{fig:traces_constant}
  \end{subfigure}
  \begin{subfigure}{0.49\columnwidth}
    \includegraphics[height=66pt]{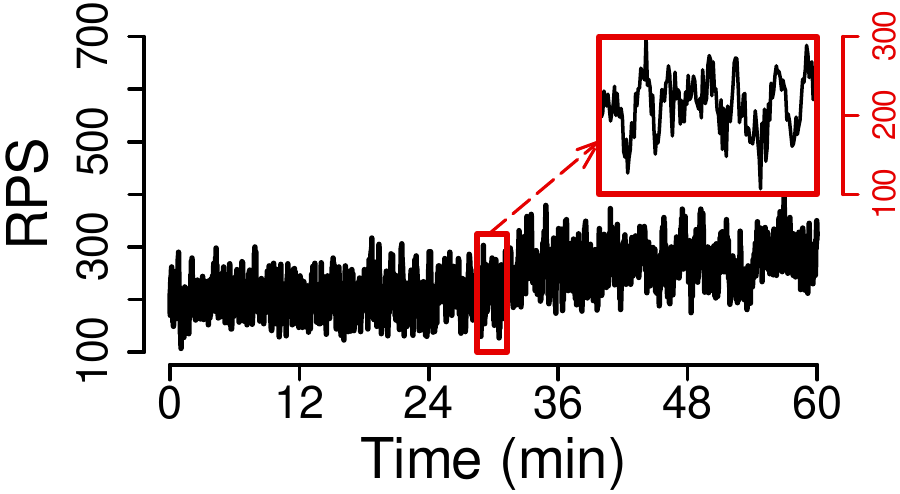}
    \caption{Noisy}
    \label{fig:traces_noisy}
  \end{subfigure}
  \hfill
  \begin{subfigure}{0.49\columnwidth}
    \includegraphics[height=66pt]{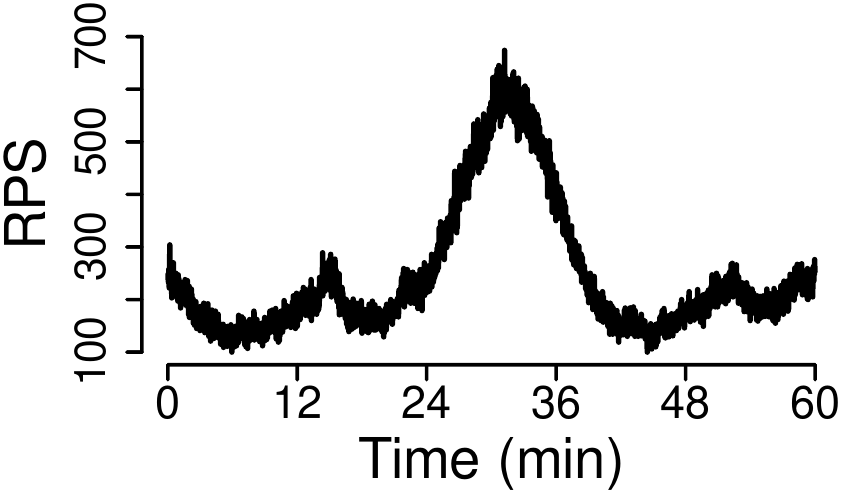}
    \caption{Bursty}
    \label{fig:traces_bursty}
  \end{subfigure}
  \caption{Our workload traces capture common patterns of RPS (requests per
  second) on an hourly basis. These patterns have been observed in real-world
  scenarios: Puffer streaming requests~\cite{puffer}, Google cluster
  usage~\cite{google_cluster_trace}, and Twitter tweets~\cite{twitter_api}.
  We also recorded a full 21-day workload trace from a global cloud provider for long-term evaluation. We scale these traces accordingly for each benchmark application to saturate the cluster (Appendix~\ref{sec:appendix_traces_rps}).}
  \label{fig:traces}
\end{figure}

\para{Experiment setup.}
We generate workloads with Locust~\cite{locust}, which is configured to mix
application requests (Appendix~\ref{sec:appendix_workload}) to stress as many services
as possible. Locust replays workload traces to reproduce RPS (requests per
second). The first set of traces captures hourly RPS patterns, which are
commonly observed in production environments: Puffer's streaming
requests~\cite{puffer}, Google's cluster usage~\cite{google_cluster_trace}, and
Twitter tweets~\cite{twitter_api}. Figure~\ref{fig:traces} illustrates these
patterns: diurnal, constant, noisy, and bursty. We also keep a full 21-day
workload trace from a global cloud provider for long-term evaluation. Depending
on the complexity of benchmark applications, we scale traces accordingly to
saturate the cluster (Appendix~\ref{sec:appendix_traces_rps}).

Each experiment ends when Locust finishes replaying a trace. For comparisons, we
record the following per-hour measurements: \textit{\textbf{(1)}} the average
number of CPU cores allocated, and \textit{\textbf{(2)}} the application end-to-end P99
latency.

Our testbeds consist of a 160-core cluster (over five 32-core Azure VMs with AMD
EPYC 7763 processors) and a 512-core cluster (over six 64-core and four 32-core
physical servers with Intel Xeon Silver 4216 processors).

\subsection{Application SLO and resource saving}
\label{sec:eval_slo_perf}

\definecolor{deepgreen}{rgb}{0.02,0.4,0.03}
\newcommand{\myperc}[1]{\scriptsize\color{deepgreen}{#1}}

\begin{table}[t]
  \begin{subtable}{1\columnwidth}
    \centering
    \footnotesize
    \setlength\tabcolsep{2pt}
    \renewcommand{\arraystretch}{1}
    \begin{tabular}{lcccc}
    \toprule
    Workload & \colorbox{green!15}{\sysname} & K8s-CPU & \colorbox{gray!15}{K8s-CPU-Fast} & Sinan \\
    \midrule
    Diurnal  & 30.4 & 58.0 \myperc{($\downarrow$47.59\%)} & 41.2 \myperc{($\downarrow$26.21\%)} & 278.4 \myperc{($\downarrow$89.08\%)} \\
    Constant & 21.7 & 24.8 \myperc{($\downarrow$12.50\%)} & 27.3 \myperc{($\downarrow$20.51\%)} & 279.9 \myperc{($\downarrow$92.25\%)} \\
    Noisy    & 15.5 & 23.6 \myperc{($\downarrow$34.32\%)} & 17.7 \myperc{($\downarrow$12.43\%)} & 251.8 \myperc{($\downarrow$93.84\%)} \\
    Bursty   & 17.7 & 27.1 \myperc{($\downarrow$34.69\%)} & 21.9 \myperc{($\downarrow$19.18\%)} & 268.3 \myperc{($\downarrow$93.40\%)} \\
    \bottomrule
    \end{tabular}
    \caption{Train-Ticket application (SLO: 1,000 ms P99 latency)}
    \label{tbl:trainticket}
  \end{subtable}
  \begin{subtable}{1\columnwidth}
    \vspace{5pt}
    \centering
    \footnotesize
    \setlength\tabcolsep{2pt}
    \renewcommand{\arraystretch}{1}
    \begin{tabular}{lcccc}
    \toprule
    Workload & \colorbox{green!15}{\sysname} & \colorbox{gray!15}{K8s-CPU} & K8s-CPU-Fast & Sinan \\
    \midrule
    Diurnal  & 77.5 & 93.9  \myperc{($\downarrow$17.47\%)} & 115.5 \myperc{($\downarrow$32.90\%)} & 162.7 \myperc{($\downarrow$52.37\%)} \\
    Constant & 88.7 & 115.6 \myperc{($\downarrow$23.27\%)} & 118.8 \myperc{($\downarrow$25.34\%)} & 149.7 \myperc{($\downarrow$40.75\%)} \\
    Noisy    & 57.5 & 66.5  \myperc{($\downarrow$13.53\%)} & 105.1 \myperc{($\downarrow$45.29\%)} & 105.2 \myperc{($\downarrow$45.34\%)} \\
    Bursty   & 50.0 & 67.5  \myperc{($\downarrow$25.93\%)} & 99.7  \myperc{($\downarrow$49.85\%)} & 111.9 \myperc{($\downarrow$55.32\%)} \\
    \bottomrule
    \end{tabular}
    \caption{Social-Network application (SLO: 200 ms P99 latency)}
    \label{tbl:social}
  \end{subtable}
  \begin{subtable}{1\columnwidth}
      \vspace{5pt}
      \centering
      \footnotesize
      \setlength\tabcolsep{2pt}
      \renewcommand{\arraystretch}{1}
      \begin{tabular}{lcccc}
      \toprule
      Workload & \colorbox{green!15}{\sysname} & K8s-CPU & \colorbox{gray!15}{K8s-CPU-Fast} & Sinan \\
      \midrule
      Diurnal  & 15.3 & 15.7 \myperc{($\downarrow$2.55\%)}  & 16.5 \myperc{($\downarrow$7.27\%)} & 45.5 \myperc{($\downarrow$66.37\%)} \\
      Constant & 11.2 & 11.5 \myperc{($\downarrow$2.61\%)}  & 11.3 \myperc{($\downarrow$0.88\%)} & 21.2 \myperc{($\downarrow$47.17\%)} \\
      Noisy    & 10.8 & 12.1 \myperc{($\downarrow$10.74\%)} & 11.6 \myperc{($\downarrow$6.90\%)} & 65.9 \myperc{($\downarrow$83.61\%)} \\
      Bursty   & 10.1 & 15.7 \myperc{($\downarrow$35.67\%)} & 10.9 \myperc{($\downarrow$7.34\%)} & 63.1 \myperc{($\downarrow$83.99\%)} \\
      \bottomrule
      \end{tabular}
      \caption{Hotel-Reservation application (SLO: 100 ms P99 latency)}
      \label{tbl:hotel}
   \end{subtable}
   \caption{Average number of CPU cores that {\sysname} and baselines
   allocate to satisfy the SLO (and thus latencies are elided).
   Percentages in parentheses quantify {\sysname}'s CPU savings over each
   baseline. The overall best-performing baseline for each application is highlighted
   in gray.  For K8s-CPU and K8s-CPU-Fast, we manually
   search for their optimal utilization thresholds (to minimize the average
   CPU allocation), per application and
   workload trace (Appendix~\ref{sec:appendix_k8s_cpu_targets}).}
   \label{tbl:resource_allocation}
\end{table}

\begin{figure}[t]
  \centering
  \includegraphics[width=0.95\columnwidth]{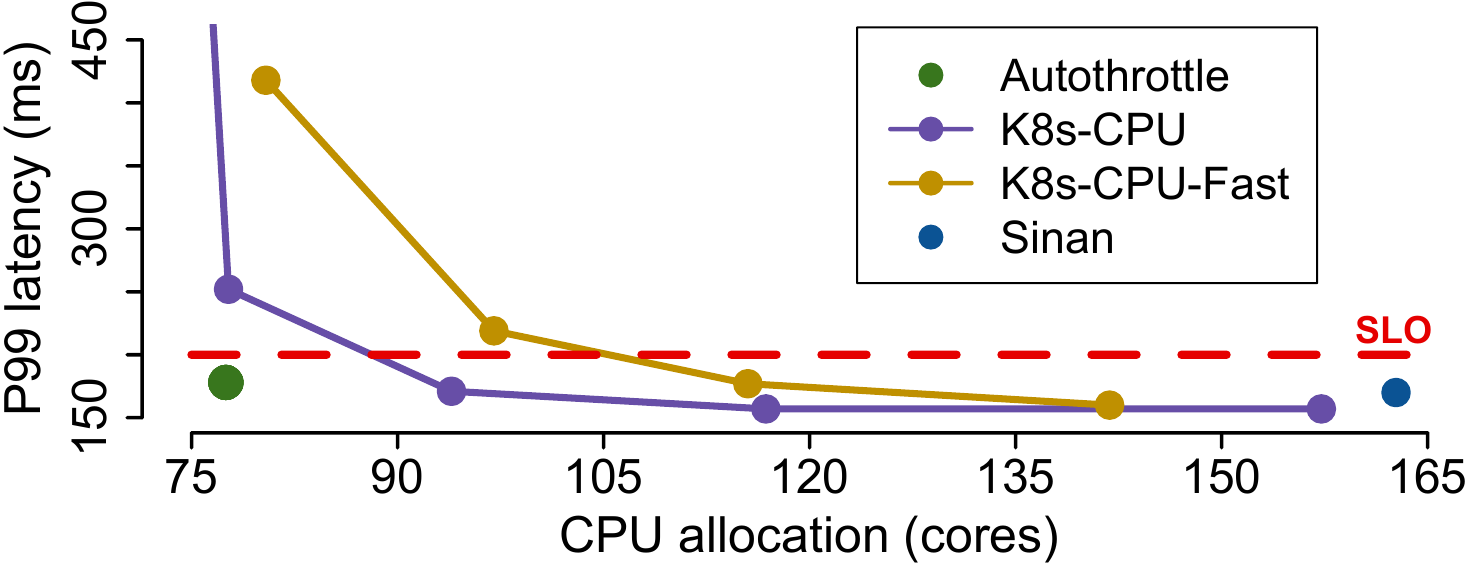}
  \caption{Application latency vs. CPU allocations, as we vary the two
  baselines' CPU utilization threshold for Social-Network under the diurnal
  workload trace. Dashed red line illustrates the 200 ms SLO.
  {\sysname} is able to maintain the SLO with the minimum CPU allocation.}
  \label{fig:social_latency_vs_allocation}
\end{figure}

We evaluate the amount of CPU resources that {\sysname} saves over baselines,
while every algorithm tries to maintain the hourly SLO over time. To ensure that
all baselines can achieve their best results, we manually identify and tune
their settings prior to experiments (Appendix~\ref{sec:appendix_k8s_cpu_targets}).

Table~\ref{tbl:resource_allocation} summarizes empirical results on the 160-core
cluster, and {\sysname} outperforms baselines in all applications. We make the
following observations, with respect to heuristic-based baselines. First, in
Social-Network, {\sysname} saves up to 25.93\% of CPU resources (or 17.5 cores)
over K8s-CPU, and up to 49.85\% of CPU resources (or 49.7 cores) over
K8s-CPU-Fast. Delving into empirical results,
Figure~\ref{fig:social_latency_vs_allocation} suggests that tuning the
baselines' CPU utilization thresholds does not make them outperform
{\sysname}. Taking the diurnal workload as an example, the figure shows that
{\sysname} is able to maintain the application SLO with the minimum CPU
allocation---{\sysname} achieves a P99 latency of 178 ms with only
77.5 cores, whereas K8s-CPU achieves 177 ms with 115.5 cores and K8s-CPU-Fast
achieves 171 ms with 93.9 cores, at best.
When allocating a comparable number of CPUs ($\sim$80 cores) to {\sysname},
K8s-CPU and K8s-CPU-Fast would violate the SLO, resulting in latencies
of 252~ms and 418~ms respectively.
Second, {\sysname} has a relatively
low resource reduction on Hotel-Reservation. This is due to the application simplicity
where requests traverse an average of only 3 microservices. A
similar observation can be made for the constant workload trace, where the
relatively static RPS pattern simplifies scaling decisions.

Furthermore, Table~\ref{tbl:resource_allocation} shows that {\sysname}
outperforms the ML-enabled baseline, Sinan. Its CPU saving is at least
40.75\% (or 61 cores for Social-Network). Deeper investigations suggest two
reasons for this gap. First, while we are able to achieve the model accuracy
published by authors (e.g., training RMSE of 22.39 and validation RMSE of 22.07,
for Social-Network) after 20+ hours of training, this non-negligible error can still
mislead scaling decisions, especially for non-constant workloads. Second, in
order to reduce training costs, Sinan learns to make relatively coarse-grained
CPU allocation adjustments (i.e., $\pm$1 core, $\pm$10\% cores, and $\pm$50\%
cores).

\begin{figure}[t]
  \centering
  \includegraphics[width=1\columnwidth]{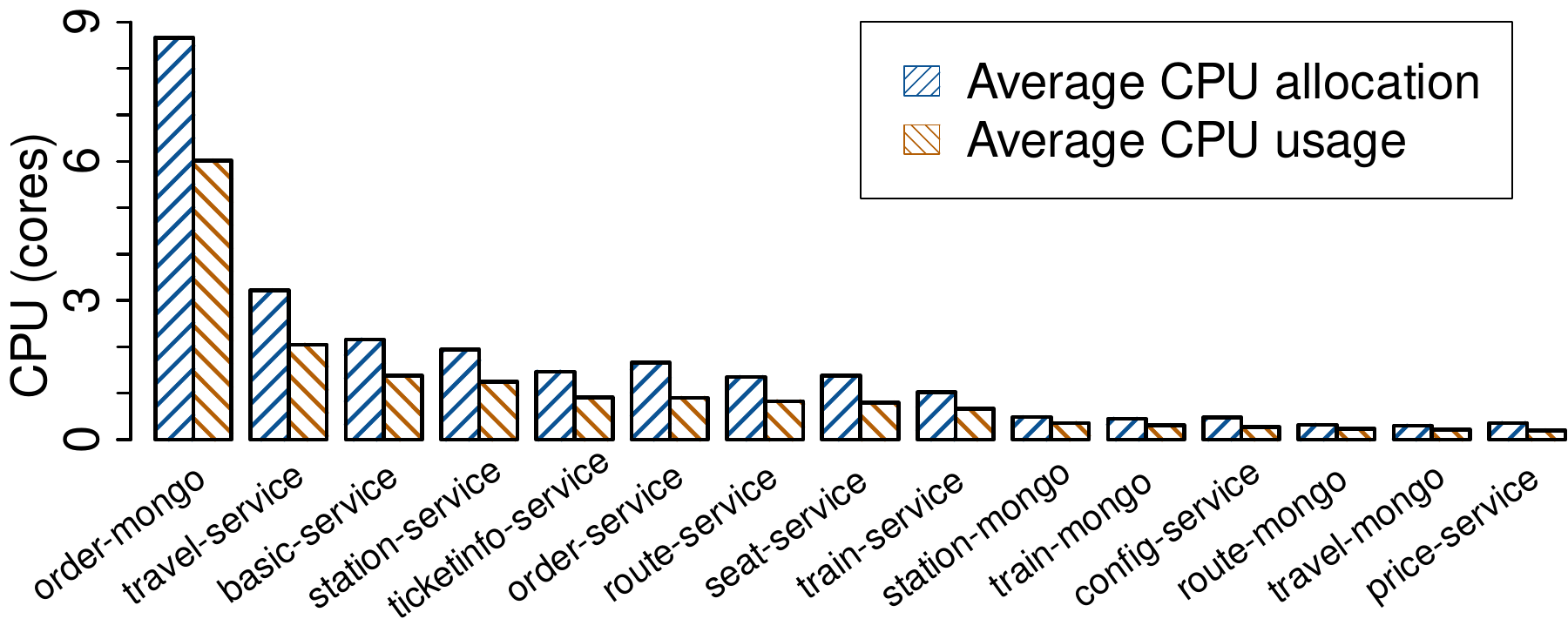}
  \caption{{\sysname} tailors CPU allocations to each microservice's resource usage.
  Figure shows top 15 microservices with the highest CPU usage in Train-Ticket
  under the diurnal workload trace.}
  \label{fig:trainticket_allocation_diurnal}
\end{figure}

\begin{figure}[t]
  \begin{subfigure}{0.496\columnwidth}
    \includegraphics[width=1\columnwidth]{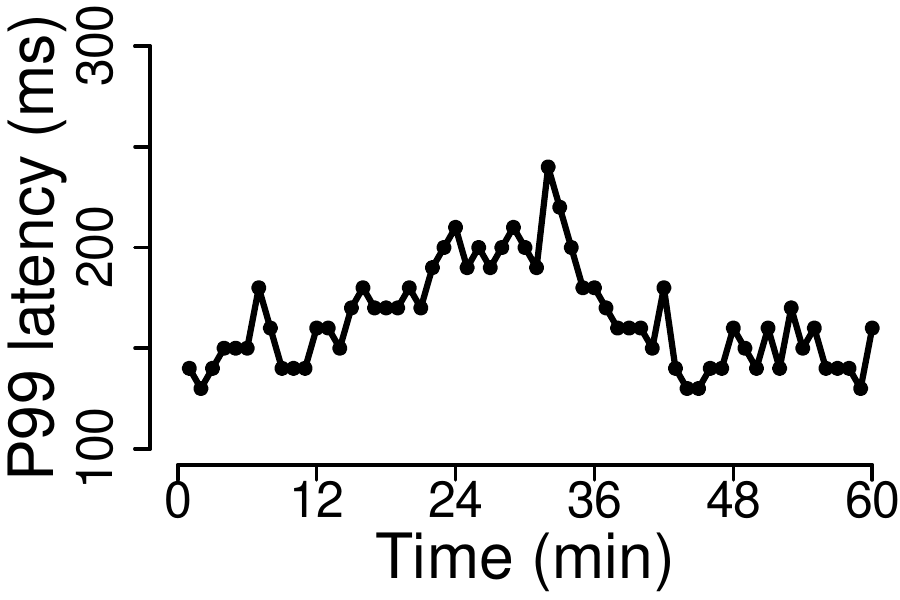}
    \vspace{-16pt}
    \caption{Application latency}
  \end{subfigure}
  \hfill
  \begin{subfigure}{0.496\columnwidth}
    \includegraphics[width=1\columnwidth]{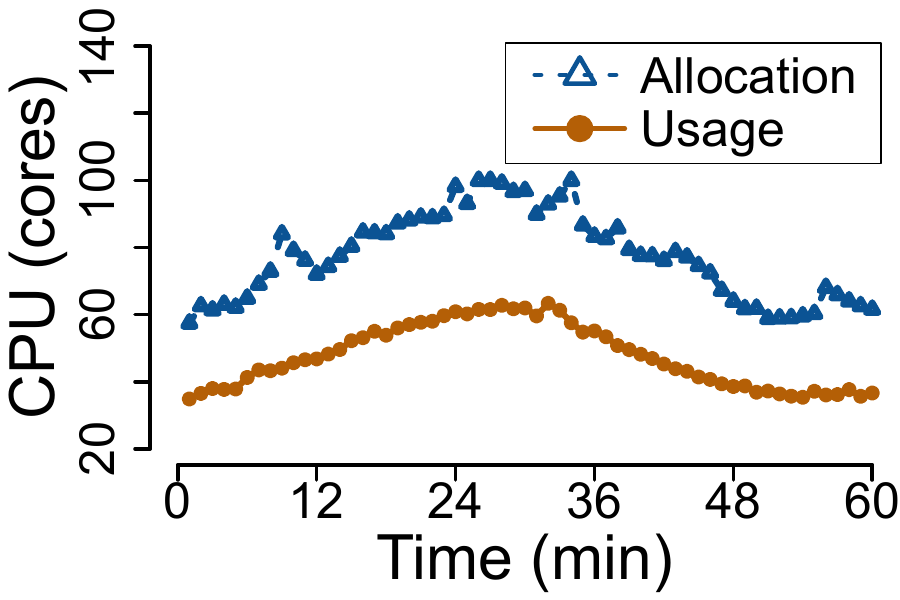}
    \vspace{-16pt}
    \caption{CPU allocation and usage}
  \end{subfigure}
  \begin{subfigure}{0.496\columnwidth}
    \vspace*{5pt}
    \includegraphics[width=1\columnwidth]{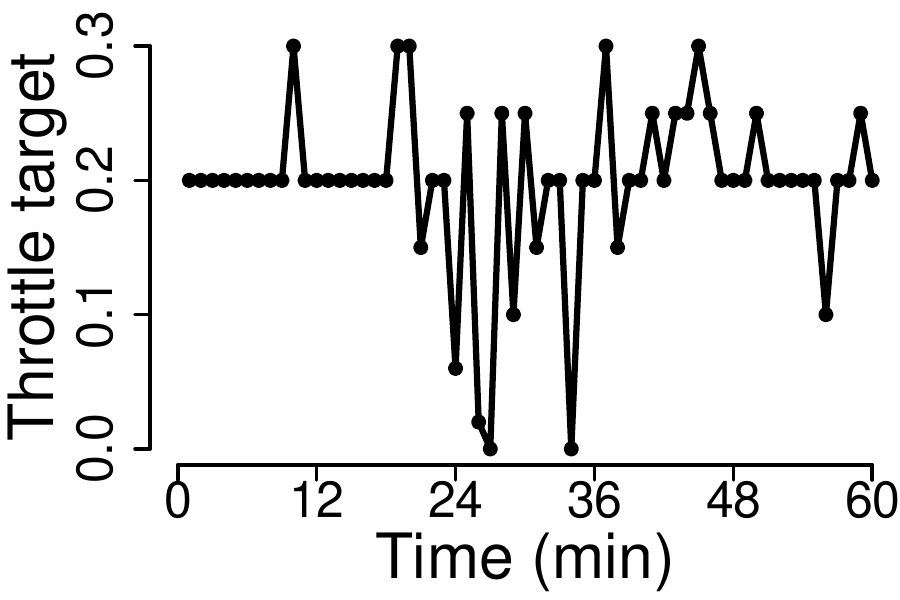}
    \vspace{-16pt}
    \caption{Throttle target \#1}
  \end{subfigure}
  \hfill
  \begin{subfigure}{0.496\columnwidth}
    \vspace*{5pt}
    \includegraphics[width=1\columnwidth]{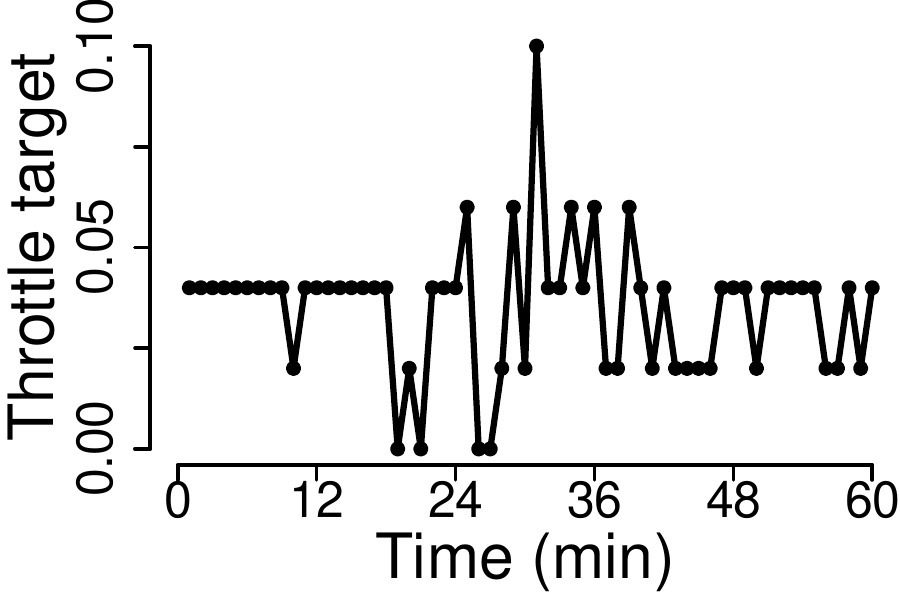}
    \vspace{-16pt}
    \caption{Throttle target \#2}
  \end{subfigure}
  \caption{Measurements of Social-Network under diurnal workload. Figures (a) and (b)
  show the latency and CPU statistics achieved by {\sysname}. Figures (c) and (d)
  demonstrate how {\sysglobal} adjusts throttle targets in response to
  time-varying workload for the two CPU usage groups (Appendix~\ref{sec:appendix_cluster}).}
  \label{fig:social_diurnal}
\end{figure}

Resource savings from Table~\ref{tbl:resource_allocation} are due to {\sysname}'s ability to tailor
CPU allocations across services and over time.
For example, Figure~\ref{fig:trainticket_allocation_diurnal} looks at top 15 microservices with
the highest CPU usage, under diurnal workload in Train-Ticket.
We note that CPU allocation is noticeably lower for services with less CPU usage
(e.g., \texttt{price-service}). Under the same workload,
Figure~\ref{fig:social_diurnal} illustrates how {\sysglobal} updates performance
targets---as the RPS varies over time, {\sysglobal} selects appropriate throttle
targets to adjust CPU allocations and maintain the P99 latency.
Note that per-minute P99 latencies are displayed in this figure, different
from the hourly P99 latencies shown in the remaining evaluation.

\subsection{Microbenchmarks}
\label{subsec:microbenchmarks}

\begin{figure}[t]
  \begin{subfigure}{0.495\columnwidth}
    \includegraphics[width=1\columnwidth]{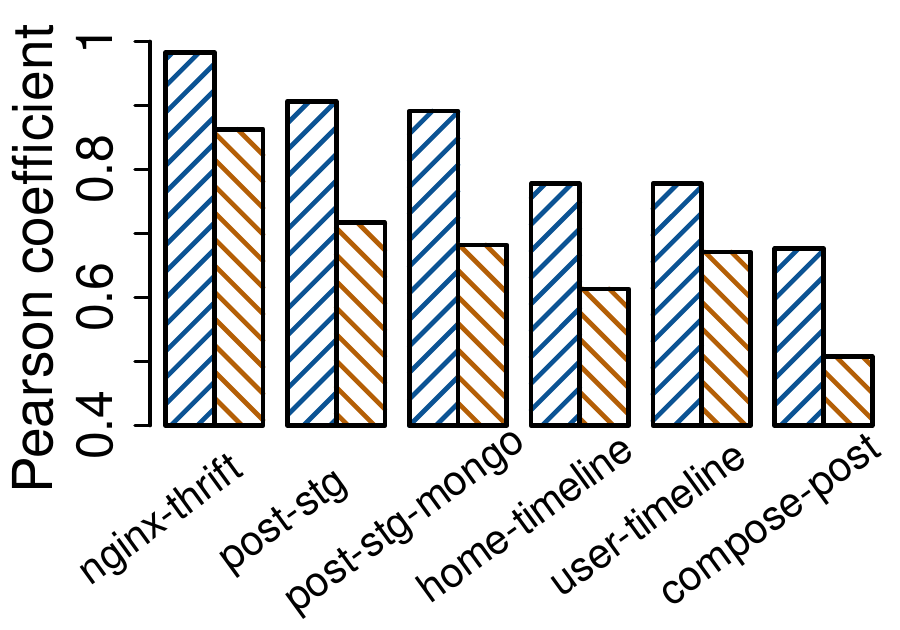}
    \vspace{-16pt}
    \caption{Social-Network}
    \label{fig:pearson_social}
  \end{subfigure}
  \hfill
  \begin{subfigure}{0.495\columnwidth}
    \includegraphics[width=1\columnwidth]{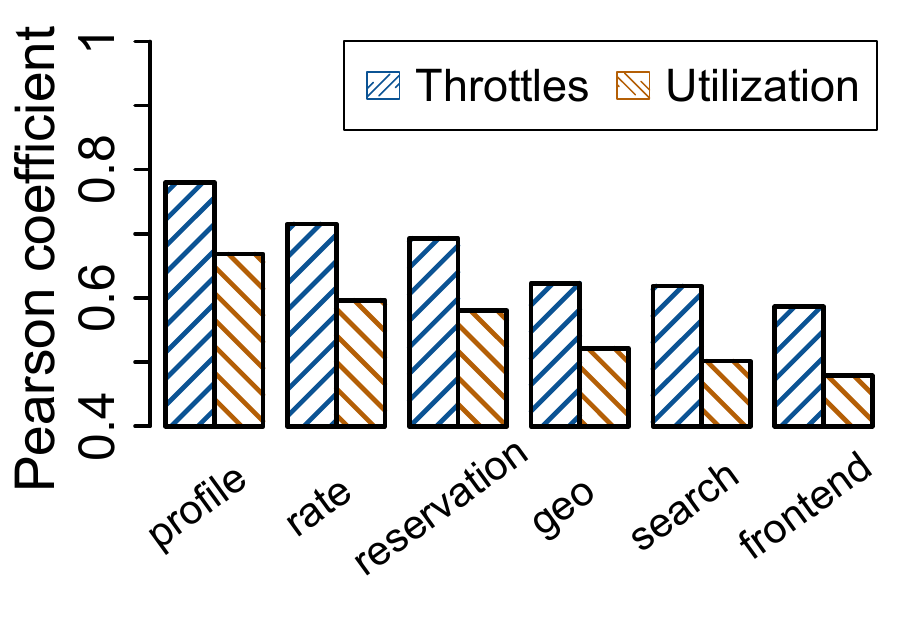}
    \vspace{-16pt}
    \caption{Hotel-Reservation}
    \label{fig:pearson_hotels}
  \end{subfigure}
  \caption{As a proxy metric, CPU throttles exhibit a higher correlation with
  application latencies than CPU utilization. The figure shows top
  microservices with highest CPU usage.}
  \label{fig:pearson}
\end{figure}

\para{Correlation of proxy metrics to latencies.}
Compared with the prevalent proxy metric for estimating resource
demand---CPU utilization,
our use of CPU throttles is motivated by the higher
correlation with application latencies as demonstrated by
Figure~\ref{fig:pearson}.
For each service in
Social-Network, we manually set its CPU quota (i.e., \texttt{cpu.cfs\_quota\_us}) to
40 uniformly distributed values. Then, we measure CPU utilization, CPU
throttle counts, and application P99 latency, at 300 RPS.
We compute the Pearson
correlation coefficient for \textit{\textbf{(1)}} latency vs. CPU
throttles, and \textit{\textbf{(2)}} latency vs. CPU utilization.
Figure~\ref{fig:pearson_social} focuses on Social-Network microservices using the
most CPU cores. In all cases, CPU throttles exhibit a higher
correlation than CPU utilization, suggesting a stronger linear
relationship. Figure~\ref{fig:pearson_hotels} shows the same conclusion for
Hotel-Reservation.

Recall that {\syslocals}
continuously collect local CPU throttles for resource control
(\S\ref{subsubsec:scale_up}), and {\sysglobal} distributes
CPU-throttle-based performance targets (\S\ref{subsubsec:tower_cb}). A high
correlation suggests that CPU throttling is indicative of the latency
and suitable to track locally in {\syslocal} as a target for maintaining the
SLO. The learning process in {\sysglobal} can also be simplified given a clear
relationship between CPU throttles and application latencies.

\begin{figure}[t]
  \begin{subfigure}{0.485\columnwidth}
    \includegraphics[width=\columnwidth]{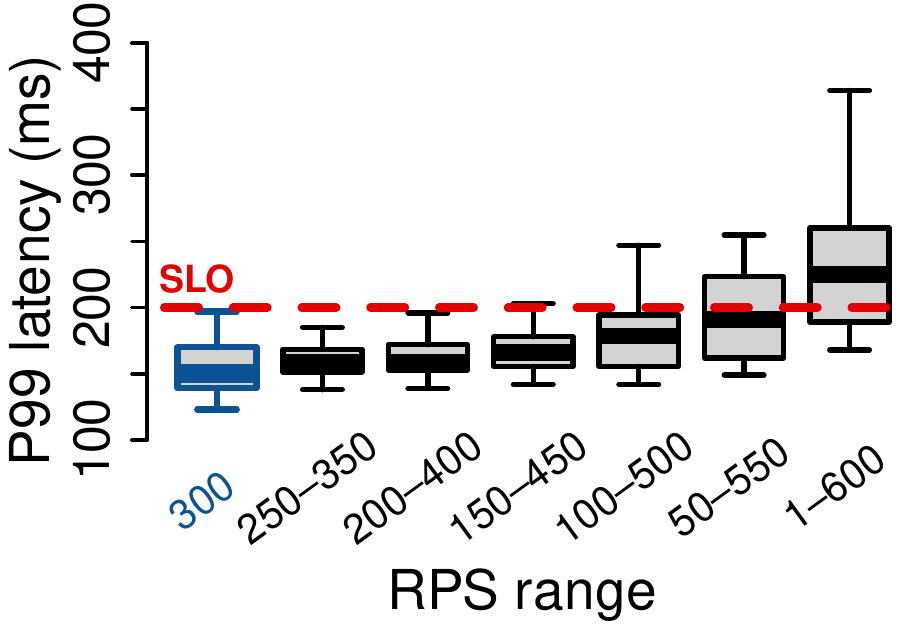}
    \vspace{-16pt}
    \caption{Social-Network}
    \label{fig:social_target_rps_sensitivity_1min}
  \end{subfigure}
  \hfill
  \begin{subfigure}{0.485\columnwidth}
    \includegraphics[width=\columnwidth]{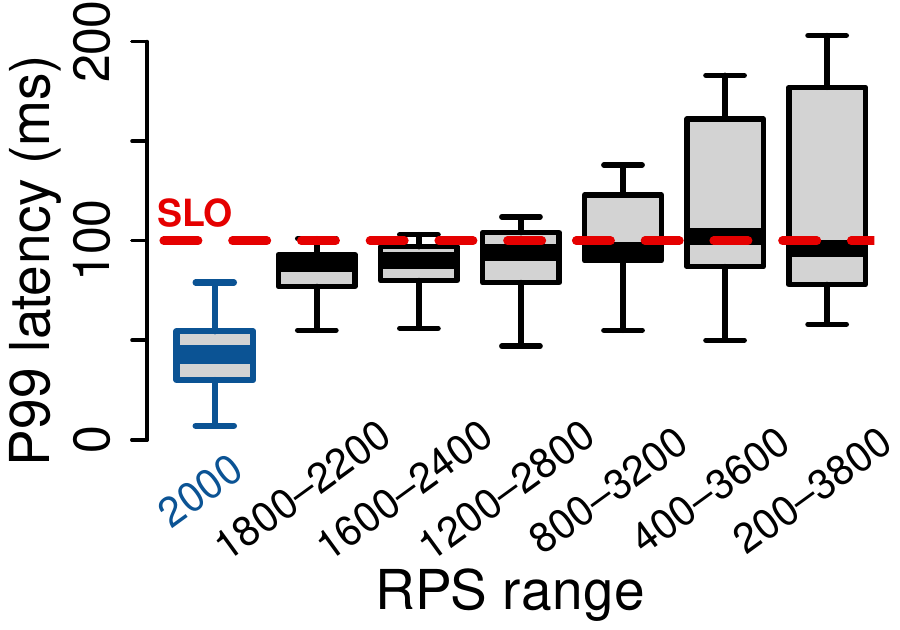}
    \vspace{-16pt}
    \caption{Hotel-Reservation}
    \label{fig:hotels_target_rps_sensitivity_1min}
  \end{subfigure}
  \caption{{\syslocal} maintains latency SLO under some workload
  fluctuations. Boxplots show latency variances, from reusing the first blue boxplot's
  performance target.}
  \label{fig:target_rps_sensitivity_1min}
\end{figure}

\para{Tolerance to short-term workload fluctuations.}
Figure~\ref{fig:target_rps_sensitivity_1min} shows
that {\syslocals} can tolerate short-term local
workload fluctuations, even with static throttle targets.
The experiment starts by finding a throttle target for Social-Network's 200 ms
SLO, at 300 RPS. Then, we reuse this target while
instrumenting Locust to fluctuate RPS in a one-minute window for 60 minutes. The
fluctuation ranges from 100 (i.e., RPS$=$250--350) to 600 (i.e., RPS$=$1--600). In
Figure~\ref{fig:social_target_rps_sensitivity_1min}, boxplots summarize the latency
variance of 60 windows. {\sysname} can keep the application P99 latency under
SLO for a fluctuation range up to 300 (i.e., RPS$=$150--450),
or up to 500 (i.e., RPS$=$50--550) if we consider the
median value instead.
Similarly, Figure~\ref{fig:hotels_target_rps_sensitivity_1min} shows RPS
fluctuation tolerance up to 800 (i.e., RPS$=$1,600--2,400) for Hotel-Reservation.

\begin{figure*}[t]
  \centering
  \begin{subfigure}{1\textwidth}
    \centering
    \includegraphics[width=\columnwidth]{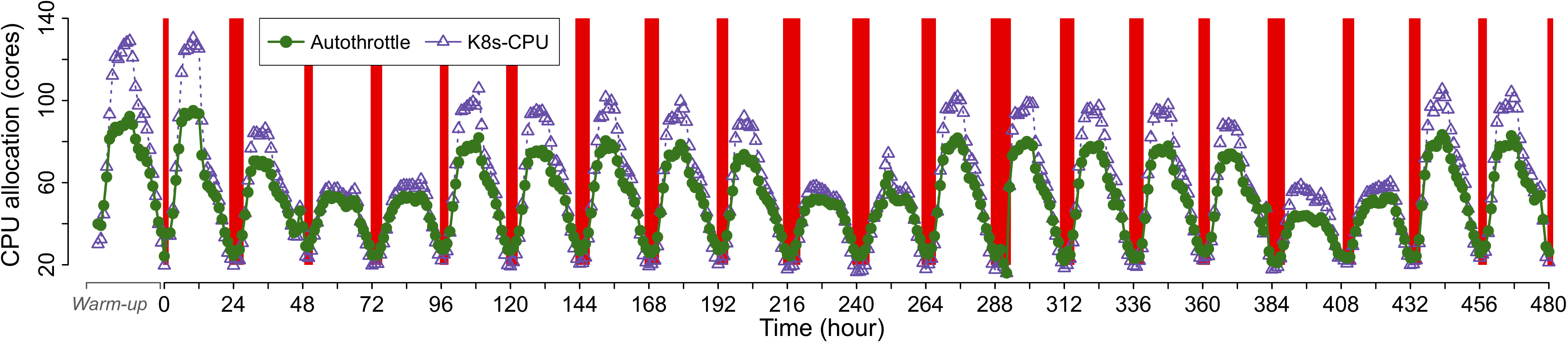}
    \vspace{-16pt}
    \caption{CPUs allocated by {\sysname} and the K8s-CPU baseline. Red boxes
    highlight hours of K8s-CPU's SLO violations.}
    \label{fig:bing-multidays-cpu}
  \end{subfigure}
  \begin{subfigure}{1\textwidth}
    \centering
    \includegraphics[width=\columnwidth]{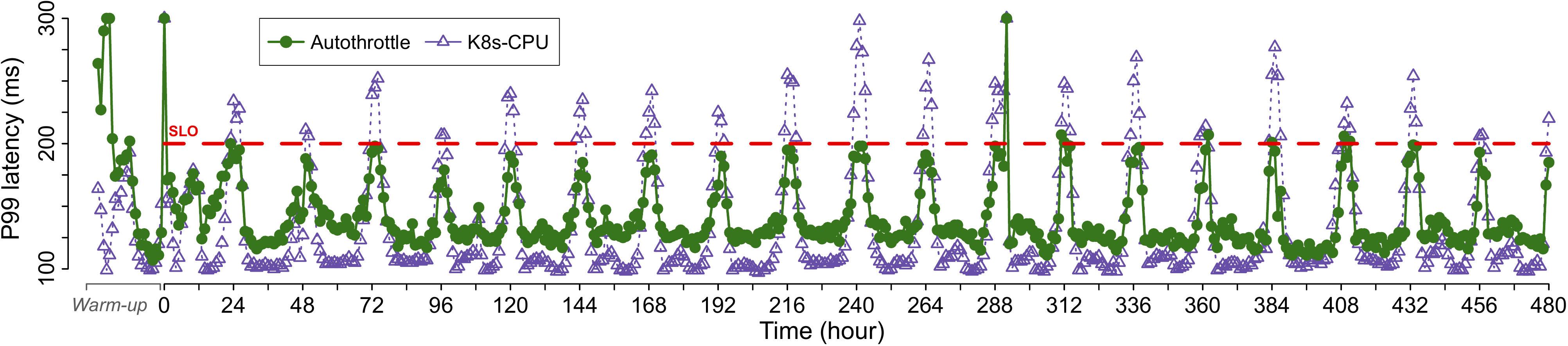}
    \vspace{-16pt}
    \caption{Social-Network's P99 latency, as achieved by {\sysname} and the K8s-CPU
    baseline. Dashed red line illustrates the 200 ms SLO.}
    \label{fig:bing-multidays-latency}
  \end{subfigure}
  \caption{A 21-day study on Social-Network with real-world workload trace from
  a global cloud provider. Compared with {\sysname}, the K8s-CPU baseline
  over-allocates an average of 12.1 and up to 35.2
  cores, and triggers 71 hourly SLO violations.}
  \label{fig:bing-multidays}
  \vspace{-5pt}
\end{figure*}

The tolerance to short-term workload fluctuations stems from the
use of performance targets (vs. exact resource allocations), which hide
service-level resource details from {\sysglobal} and enable {\syslocal}s to
autonomously adjust resource allocations. This tolerance is vital as
it frees {\sysglobal} from the excessive recomputation of
performance targets (\S\ref{subsubsec:tower_cb}).

\para{Number of performance targets.}
Rather than generating separate performance targets for individual microservices,
{\sysglobal} clusters microservices into two categories based on their average
CPU usage, reducing the action space to two targets
(\S\ref{subsubsec:tower_cb}).
To assess this design, we empirically compare the
performance of 1, 2, 3, and 4 targets,
under the constant workload trace.
In each scenario, we manually search for the best-performing set of throttle targets
that satisfy the SLO using the minimum number of CPU cores.
For Social-Network, {\sysname} allocates 70.8,
55.9, 55.1, and 54.7
cores with 1 to 4 targets, respectively.
Hotel-Reservation consistently uses the largest target (0.3) to meet the
SLO on this trace, regardless of the number of targets.
For Train-Ticket, the allocation is 18.6,
18.1, and 18.1 cores with 1 to 3 targets (exhaustive search is infeasible for 4 targets).
Overall, these results suggest a diminishing return beyond 2 targets.

\para{Load-stressing to the limit.}
We stress resource managers, by pushing Locust's RPS to the application's upper
limit. This is the breaking point (before application crashing) when almost all
CPU cores are allocated. To this end, we stress Social-Network at constant RPS of 600
and 700, on the 160-core cluster. At 600 RPS,
{\sysname} still achieves a CPU core saving of 27.67\% and better
SLO---it achieves a P99 latency of 202~ms with only 98.3 cores, whereas K8s-CPU
achieves 216~ms with 135.9 cores and K8s-CPU-Fast achieves 235~ms with 133.1
cores. Finally, at 700 RPS, {\sysname} achieves a P99 latency of 452~ms with only 106.8
cores, whereas K8s-CPU achieves 600~ms with 153.1 cores, and K8s-CPU-Fast
achieves 551~ms with 143.8 cores.

\para{Ablation study for contextual bandits.}
We investigate two aspects that can impact {\sysglobal}'s contextual bandits.
The first is the number of available throttle targets to choose from in the
action space (\S\ref{subsubsec:tower_cb}). For the constant workload trace,
reducing from 9 to 4 throttle targets
results in over-allocating 5.6 CPU cores (or 10.03\%) for Social-Network, and 0.7 CPU
cores (or 3.49\%) for Train-Ticket. The second is the use of neural networks
(\S\ref{sec:impl}). Under various workload patterns on Social-Network, we test
a linear model and neural networks with different numbers of hidden units, but
their difference in CPU allocation is small. None of the tested models violates
the SLO. We include the results in Appendix~\ref{sec:appendix_vw}.

\subsection{Long-term evaluation}
\label{sec:long_term}

We perform a 21-day study with real-world workload trace from a global cloud
provider. Experiments are performed with Social-Network on the 160-core cluster, and
an hourly SLO of 200~ms is set on P99 latency. We compare {\sysname} with
K8s-CPU, the best-performing baseline from \S\ref{sec:eval_slo_perf}. We use day
1 for training and tuning {\sysname} and K8s-CPU. For the former, we train the
{\sysglobal}'s model. For the latter, we spend 24 man-hours to manually identify its
best CPU utilization threshold.

Figure~\ref{fig:bing-multidays} illustrates the results over the entire period.
Figure~\ref{fig:bing-multidays-cpu} shows the CPU core saving that {\sysname}
achieves every hour, over the K8s-CPU baseline. First, {\sysname} can save up to
35.2 cores (or an average saving of 12.1 cores) over K8s-CPU. Second, although there
are days when K8s-CPU allocates fewer CPUs (e.g., an hourly
average of \textminus2.77 CPU cores on day 4), these are also the days when K8s-CPU has a
high chance of triggering SLO violations. In total, K8s-CPU
violates the hourly SLO 71 times (highlighted by red boxes in
Figure~\ref{fig:bing-multidays-cpu}). On the other hand, {\sysname} reduces
SLO violations to only 5 times---an investigation reveals that these hours'
workloads appear anomalous (i.e., recorded RPS jumps between 0 and
$\sim$400) and unforeseen.

Figure~\ref{fig:bing-multidays-latency} shows Social-Network's P99 latency
per hour. One observation is that {\sysname} is able to continuously maintain a
P99 latency closer to the 200 ms SLO. Since its P99 latency exhibits a much lower
variance over time, this results in a more stable application performance.

\subsection{Large-scale evaluation}
\label{sec:large_scale}

\begin{figure}[t]
    \centering
    \includegraphics[width=\columnwidth]{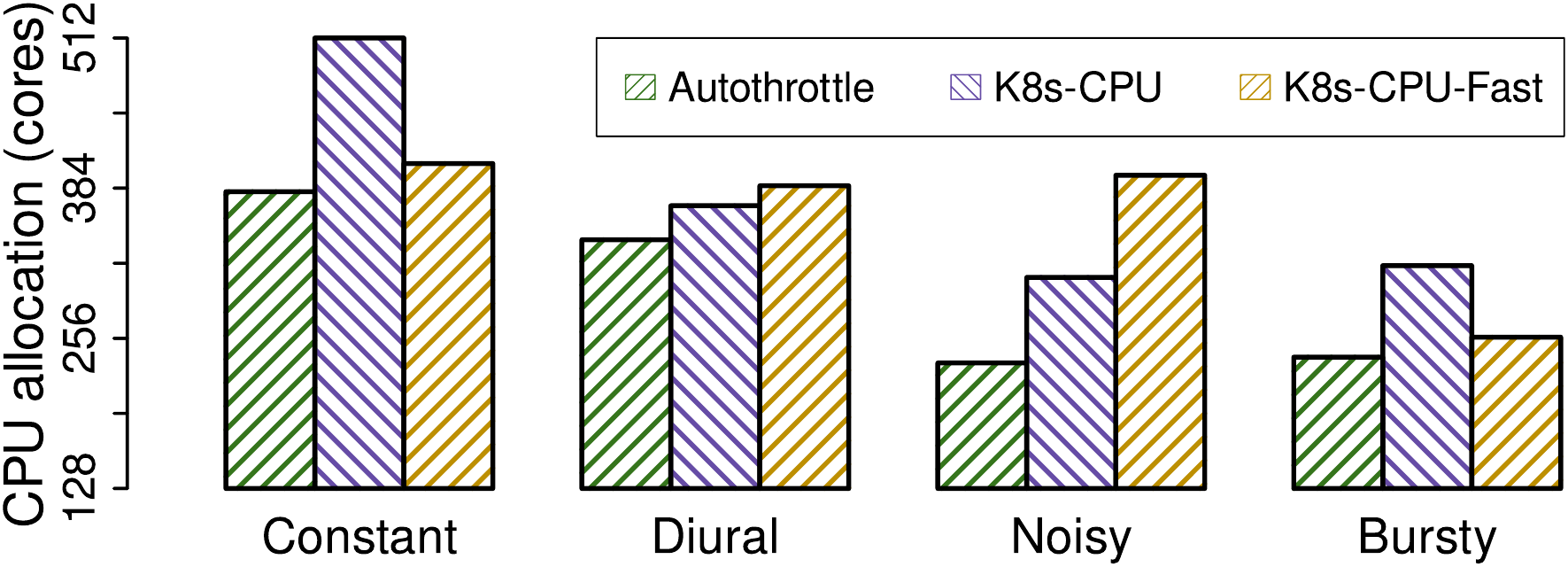}
    \caption{Number of CPU cores that {\sysname} and baselines
    allocate, to satisfy Social-Network's P99 SLO. Figure shows
    {\sysname}'s scalability on the larger 512-core cluster.}
    \label{fig:512cores_usage_allocation}
\end{figure}

We now show {\sysname}'s scalability on the larger 512-core cluster. It allows
us to push RPS beyond the breaking point of the 160-core cluster
(\S\ref{subsec:microbenchmarks}), up to 1,200 on Social-Network
(the upper limit for comparison baselines). To fully allocate all cores,
we replicate Social-Network's CPU-intensive microservices: Nginx ($\times$3) and
ML-based image classifier ($\times$6).

Figure~\ref{fig:512cores_usage_allocation} shows that {\sysname} is able to
allocate fewer CPU cores while meeting Social-Network's 200~ms P99 SLO.
Compared to the best-performing baselines, K8s-CPU and
K8s-CPU-fast, {\sysname} saves up to 28.24\% (or 150 CPU cores) and at least
5.92\% (or 24 CPU cores). Finally, we note that K8s-CPU-Fast can have a higher
CPU allocation than K8s-CPU, especially for the noisy workload trace. Since
K8s-CPU-Fast is more sensitive to CPU utilization changes than K8s-CPU,
it can sometimes accidentally scale down and lead to SLO violations. As a result,
conservatively setting K8s-CPU-Fast results in the trade-off of higher CPU allocation.

\tightsection{Related Work}
\label{sec:relwork}

\para{Cloud resource management.}
Cloud vendors have long
offered services that enable elastic scaling of VMs and their associated
resources according to user-defined
rules~\cite{azure-autoscale,aws-autoscale,gce-autoscale}.
In addition to rule-based scaling,
researchers have proposed predictive scaling,
which involves forecasting
future demand and adjusting resource allocation in advance of
any demand changes~\cite{aws-predictive-scaling,gong2010press,shen2011cloudscale,nguyen2013agile}.
Despite the cost effectiveness of these mechanisms in meeting SLOs,
they are primarily designed for VMs
(e.g., targeting monolithic
applications or relying on VM-specific techniques such as live migration),
and cannot be directly applied to microservices.
Other cluster management frameworks~\cite{mesos,omega,yarn,borg,tarcil}
that schedule jobs to clusters may be used in conjunction with {\sysname}.

\para{Vertical scaling of microservices.}
Vertical autoscalers adjust the resource limits
in a fine-grained manner, e.g., milli-cores.
Kubernetes Vertical Pod Autoscaler (VPA)~\cite{k8s_vpa} heuristically adjusts resource limits
to maintain a user-specified utilization threshold.
Autopilot~\cite{autopilot} focuses on vertical scaling,
selecting resource limits based on moving windows of historical usage
and an ML technique akin to multi-armed bandit.
Sinan~\cite{sinan} trains ML models to infer the likelihood of SLO violations
given a set of proposed CPU limits.
FIRM~\cite{firm} reacts to SLO violations and pinpoints a microservice as the
root cause, using reinforcement learning to scale up the service.
A recent work~\cite{lee2022autothrottle} (also named ``Autothrottle'') adjusts
the CPU quota of containers using closed-loop control to satisfy their
individual network SLOs (e.g., throughput), rather than application latency
SLOs.
{\sysname} differs from these approaches with its bi-level design and
the use of CPU throttle targets.

\para{Horizontal and hybrid scaling of microservices.}
Horizontal autoscalers operate at a
coarse-grained level by adjusting the number of replicas of a microservice.
Kubernetes Horizontal Pod Autoscaling (HPA)~\cite{k8s_autoscale} employs a mechanism
similar to VPA at its core, except for choosing the appropriate number of pods
to meet an input utilization threshold.
GRAF~\cite{graf} leverages graph neural networks to model
service dependencies.
COLA~\cite{cola}
uses a multi-armed bandit to collectively determine
the number of replicas for each microservice.
In addition, there are hybrid autoscalers that combine vertical and
horizontal scaling and apply them selectively~\cite{atom,hyscale}.
{\sysname} focuses on vertical scaling due to its fine-grained and
rapid reaction that empowers per-service controllers. As future work, we plan to
explore the integration of horizontal scaling with {\sysname}.

\para{Proxy metrics for estimating resource demand.}
In comparison to CPU throttles,
alternative service-level proxy metrics fall short in maintaining end-to-end
latency under workload changes.
Kubernetes defaults to CPU utilization~\cite{k8s_autoscale},
but high CPU usage does not always indicate an issue if
requests can still complete within the SLO~\cite{cola,ibmblog}.
Queue length~\cite{powerchief}, the number of requests pending at a
service, overlooks the complexity of individual requests,
while queuing delay~\cite{wechat_overload,breakwater} depends on the service's
threading model~\cite{mutune} and may require manual instrumentation of
each service.
The scattered nature of queues across the application, OS, and network,
further complicates precise measurement of queue length or queuing delay~\cite{sinan}.
Regardless, dynamically adapting the thresholds for these metrics
may require an application-level controller as proposed by \sysname.

\tightsection{Conclusion}
\label{sec:conclusion}

{\sysname} is a bi-level learning-assisted resource management framework for
SLO-targeted microservices.
It decouples mechanisms of SLO feedback and resource control,
and bridges them through CPU-throttle-based performance targets.
Going forward, we are extending to additional resource types such as memory and storage,
and exploring integrations with additional scaling strategies such as horizontal scaling.

\vspace{-5pt}
\section*{Acknowledgments}

We are grateful to Yang Yue and Jiayi Mao for their contributions in the early stage of this work.
We thank Neeraja J. Yadwadkar for shepherding our paper and the anonymous reviewers for their helpful comments and feedback.
We also thank Haidong Wang, Chuanjie Liu, Qianxi Zhang, Yawen Wang, Yu Gan, Fan Yang, Mao Yang,
Lidong Zhou, and Victor Bahl for their support or insightful discussions.

\newpage

\bibliographystyle{plain}
\bibliography{references}

\clearpage
\appendix

\section*{Appendices}
\label{sec:appendix}

\section{Application workload details}
\label{sec:appendix_workload}

We present the workload composition used in our
experiments. Our workload generator, Locust, follows the ratios specified
below when generating requests at a given RPS:
\vspace{6pt}

\para{Train-Ticket:}
\vspace{-8pt}
\begin{multicols}{2}
\begin{itemize}[itemsep=-3pt,topsep=0pt,leftmargin=*]
  \item \texttt{Mainpage}: 29.41\%
  \item \texttt{Travel}: 58.82\%
  \item \texttt{Assurance}: 2.94\%
  \item \texttt{Food}: 2.94\%
  \item \texttt{Contact}: 2.94\%
  \item \texttt{Preserve}: 2.94\%
\end{itemize}
\end{multicols}

\vspace{-7pt}

\para{Hotel-Reservation:}
\vspace{-8pt}
\begin{multicols}{2}
\begin{itemize}[itemsep=-3pt,topsep=0pt,leftmargin=*]
  \item \texttt{Search}: 60\%
  \item \texttt{Recommend}: 39\%
  \item \texttt{Reserve}: 0.5\%
  \item \texttt{Login}: 0.5\%
\end{itemize}
\end{multicols}

\vspace{-7pt}

\para{Social-Network:}
\vspace{4pt}
\begin{itemize}[itemsep=1pt,topsep=0pt,leftmargin=*]
  \noindent
  \begin{minipage}[c]{0.55\columnwidth}
  \item \texttt{Read-home-timeline}: 65\%
  \item \texttt{Read-user-timeline}: 15\%
  \end{minipage}
  \begin{minipage}[c]{0.4\columnwidth}
  \item \texttt{Compose-post}: 20\%
  \item[] ~
  \end{minipage}
\end{itemize}

\section{Vowpal Wabbit usage}
\label{sec:appendix_vw}

The following VW parameters are used in our experiments.
\begin{itemize}[itemsep=1pt,topsep=3pt,leftmargin=*]
  \item The doubly robust estimator~\cite{dudik2011doubly} is employed
  for policy evaluation: \texttt{---cb\_type dr}
  \item Number of available actions: \texttt{---cb\_explore 81}
  \item The native $\epsilon$-greedy is disabled to implement our customized exploration strategy (\S\ref{subsubsec:tower_cb}): \texttt{---epsilon 0}
  \item Number of hidden units in the neural network: \texttt{---nn 3}
  \item Learning rate: \texttt{-l 0.5}
\end{itemize}

\begin{figure}[t]
  \centering
  \includegraphics[width=0.7\columnwidth]{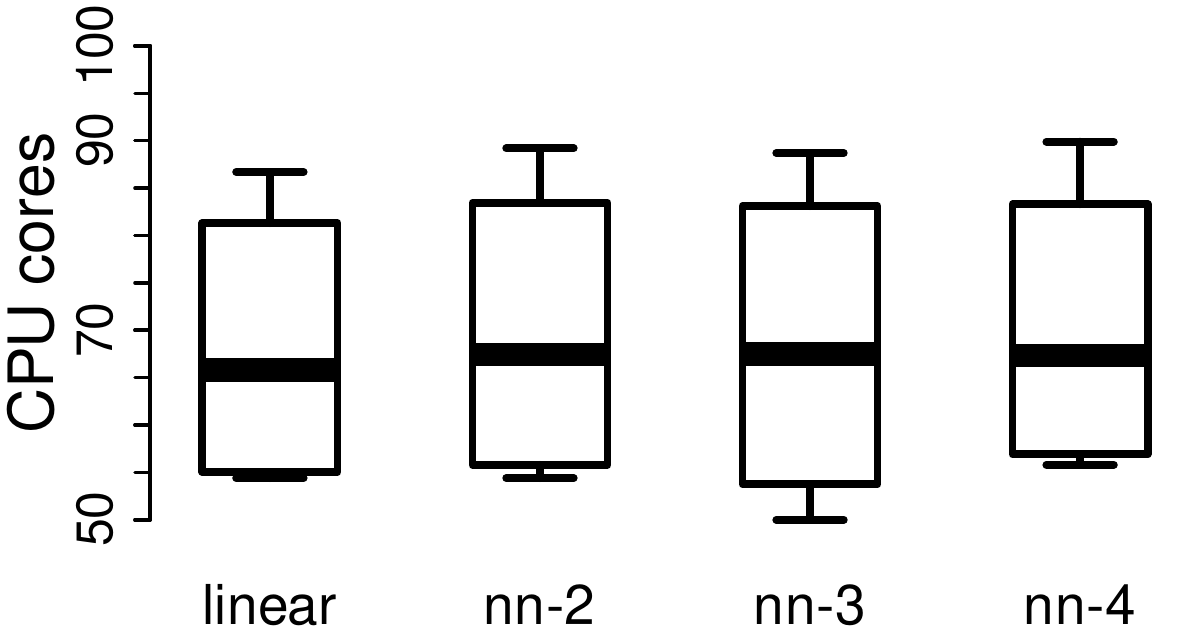}
  \caption{Different VW models---a linear model and neural networks with 2, 3,
and 4 hidden units---perform similarly on Social-Network under various
workloads (Figure~\ref{fig:traces}).}
  \label{fig:vw_linear_vs_nn_models}
\end{figure}

We also compare different VW models---a linear model and neural networks with 2,
3, 4 and hidden units, on Social-Network under the same workload patterns
(Figure~\ref{fig:traces}). Figure~\ref{fig:vw_linear_vs_nn_models} shows that
{\sysname} perform similarly across different models. We select the neural
network model with 3 hidden units, as it performs slightly better on the bursty
workload (as indicated by the lower whiskers in boxplots).

\section{Microservice clustering}
\label{sec:appendix_cluster}

{\sysglobal} clusters microservices into two groups based on their average CPU
usage, denoted ``High'' and ``Low'', using a standard $k$-means
clustering algorithm (\S\ref{subsubsec:tower_cb}).
Table~\ref{high-low-cpu-groups} presents
a breakdown of the number of services in each group.

\section{Microservice replicas}
\label{sec:appendix_deployment}

\begin{table}
  \centering
  \footnotesize
  \begin{tabular}{lcc}
  \toprule
  Application & ``High'' group & ``Low'' group \\
  \midrule
  Train-Ticket & 8 & 60 \\
  Hotel-Reservation & 6 & 11 \\
  Social-Network (160-core cluster)  &  1 & 27 \\
  Social-Network (512-core cluster)  &  2 & 26 \\
  \bottomrule
  \end{tabular}
  \caption{Number of services in each application assigned to
  the ``High'' and ``Low'' CPU usage groups.}
  \label{high-low-cpu-groups}
\end{table}

Train-Ticket and Hotel-Reservation deploy each service with one replica.
For Social-Network, we employ three replicas of \texttt{media-filter-service}
except in the large-scale evaluation (\S\ref{sec:large_scale}).
In \S\ref{sec:large_scale}, 6 replicas of
\texttt{media-filter-service} and 3 replicas of \texttt{nginx-thrift}
are employed.

\section{RPS range of workload traces}
\label{sec:appendix_traces_rps}

We scale the traces presented in Figure~\ref{fig:traces} to
saturate the cluster for each application, as documented in
Table~\ref{tbl:traces-rps}.

\begin{table}[t]
  \begin{subtable}{1\columnwidth}
    \centering
    \footnotesize
    \begin{tabular}{lccc}
    \toprule
    Workload & Min RPS & Average RPS & Max RPS \\
    \midrule
    Diurnal  & 145 & 262 & 411 \\
    Constant & 152 & 200 & 252 \\
    Noisy    &  75 & 157 & 252 \\
    Bursty   &  62 & 163 & 442 \\
    \bottomrule
    \end{tabular}
    \caption{Train-Ticket}
    \vspace{5pt}
  \end{subtable}
  \begin{subtable}{1\columnwidth}
    \centering
    \footnotesize
    \begin{tabular}{lccc}
    \toprule
    Workload & Min RPS & Average RPS & Max RPS \\
    \midrule
    Diurnal  & 1721 & 2627 & 4003 \\
    Constant & 1855 & 2002 & 2183 \\
    Noisy    &  793 & 1575 & 2470 \\
    Bursty   &  768 & 1633 & 4037 \\
    \bottomrule
    \end{tabular}
    \caption{Hotel-Reservation}
    \vspace{5pt}
  \end{subtable}
  \begin{subtable}{1\columnwidth}
    \centering
    \footnotesize
    \begin{tabular}{lccc}
    \toprule
    Workload & Min RPS & Average RPS & Max RPS \\
    \midrule
    Diurnal  & 227 & 394 & 656 \\
    Constant & 390 & 500 & 588 \\
    Noisy    & 105 & 236 & 390 \\
    Bursty   & 104 & 245 & 648 \\
    Long-term (\S\ref{sec:long_term}) & 1 & 230 & 592 \\
    \bottomrule
    \end{tabular}
    \caption{Social-Network}
    \vspace{5pt}
  \end{subtable}
  \begin{subtable}{1\columnwidth}
    \centering
    \footnotesize
    \begin{tabular}{lccc}
    \toprule
    Workload & Min RPS & Average RPS & Max RPS \\
    \midrule
    Diurnal  & 479 &  787 & 1214 \\
    Constant & 882 & 1001 & 1131 \\
    Noisy    & 232 &  472 &  771 \\
    Bursty   & 205 &  489 & 1266 \\
    \bottomrule
    \end{tabular}
    \caption{Social-Network, large-scale evaluation (\S\ref{sec:large_scale})}
    \vspace{5pt}
  \end{subtable}
  \caption{The RPS range of workload traces after being scaled to saturate the
  cluster for each application.}
  \label{tbl:traces-rps}
  \vspace{-5pt}
\end{table}

\begin{table}[t]
  \begin{subtable}{1\columnwidth}
    \centering
    \footnotesize
    \begin{tabular}{lcc}
    \toprule
    Workload & K8s-CPU & K8s-CPU-Fast \\
    \midrule
    Diurnal  & 0.4 & 0.6 \\
    Constant & 0.6 & 0.6 \\
    Noisy    & 0.5 & 0.7 \\
    Bursty   & 0.5 & 0.6 \\
    \bottomrule
    \end{tabular}
    \caption{Train-Ticket}
    \vspace{5pt}
  \end{subtable}
  \begin{subtable}{1\columnwidth}
    \centering
    \footnotesize
    \begin{tabular}{lcc}
    \toprule
    Workload & K8s-CPU & K8s-CPU-Fast \\
    \midrule
    Diurnal  & 0.7 & 0.7 \\
    Constant & 0.7 & 0.8 \\
    Noisy    & 0.6 & 0.7 \\
    Bursty   & 0.5 & 0.7 \\
    \bottomrule
    \end{tabular}
    \caption{Hotel-Reservation}
    \vspace{5pt}
  \end{subtable}
  \begin{subtable}{1\columnwidth}
    \centering
    \footnotesize
    \begin{tabular}{lcc}
    \toprule
    Workload & K8s-CPU & K8s-CPU-Fast \\
    \midrule
    Diurnal  & 0.5 & 0.5 \\
    Constant & 0.5 & 0.6 \\
    Noisy    & 0.5 & 0.4 \\
    Bursty   & 0.5 & 0.4 \\
    Long-term (\S\ref{sec:long_term}) & 0.5 & -- \\
    \bottomrule
    \end{tabular}
    \caption{Social-Network}
    \vspace{5pt}
  \end{subtable}
  \begin{subtable}{1\columnwidth}
    \centering
    \footnotesize
    \begin{tabular}{lcc}
    \toprule
    Workload & K8s-CPU & K8s-CPU-Fast \\
    \midrule
    Diurnal  & 0.6 & 0.7 \\
    Constant & 0.5 & 0.8 \\
    Noisy    & 0.5 & 0.5 \\
    Bursty   & 0.5 & 0.7 \\
    \bottomrule
    \end{tabular}
    \caption{Social-Network, large-scale evaluation (\S\ref{sec:large_scale})}
    \vspace{5pt}
  \end{subtable}
  \caption{The best-performing CPU utilization thresholds for comparison baselines, per application
  and workload trace.}
  \label{tbl:k8s-cpu-targets}
\end{table}

\section{CPU utilization thresholds in K8s-CPU and K8s-CPU-Fast}
\label{sec:appendix_k8s_cpu_targets}

In evaluating K8s-CPU and K8s-CPU-Fast, we test and select the best-performing
CPU utilization threshold from the set $\{0.1, 0.2, \dots, 0.9\}$,
for each application and each workload trace.
The selected thresholds are presented in Table~\ref{tbl:k8s-cpu-targets}.

\section{Evaluation methodology details}
\label{sec:appendix_methodology}

All experiments are performed using one-hour workload traces.
Prior to testing, certain applications require additional preparations. We warm
up Hotel-Reservation by sending 200 requests per second for 15 seconds and
waiting for 60 seconds. For Social-Network, we populate the database with 962
users, 18,812 edges, and 20,000 posts. We then warm up for 3 minutes by
incrementally increasing the RPS by 10\% every 5 seconds, up to the
initial RPS in the one-hour trace. The warm-up phase is excluded from the
calculation of P99 latency and resource allocation.

{\sysname} shown in Table~\ref{tbl:resource_allocation} is warmed up for 12
hours. The first 6 hours are the (random) exploration stage, followed by 6 hours
of normal learning with $\epsilon$$=$0.5. We use a separate 1-hour diurnal trace, which
is different from the one used for testing but has the same RPS range.
Warm up involves running 12 repetitions of this trace. During testing,
exploration is turned off completely (with
$\epsilon$ set to 0). For
Hotel-Reservation, RPS is grouped into bins of 200 due to its high RPS, while
other applications use the default bin size of 20.

\section{{\syslocal} performance}
\label{sec:appendix_captain_perf}

\begin{figure}[t]
  \begin{subfigure}{0.496\columnwidth}
    \includegraphics[width=1\columnwidth]{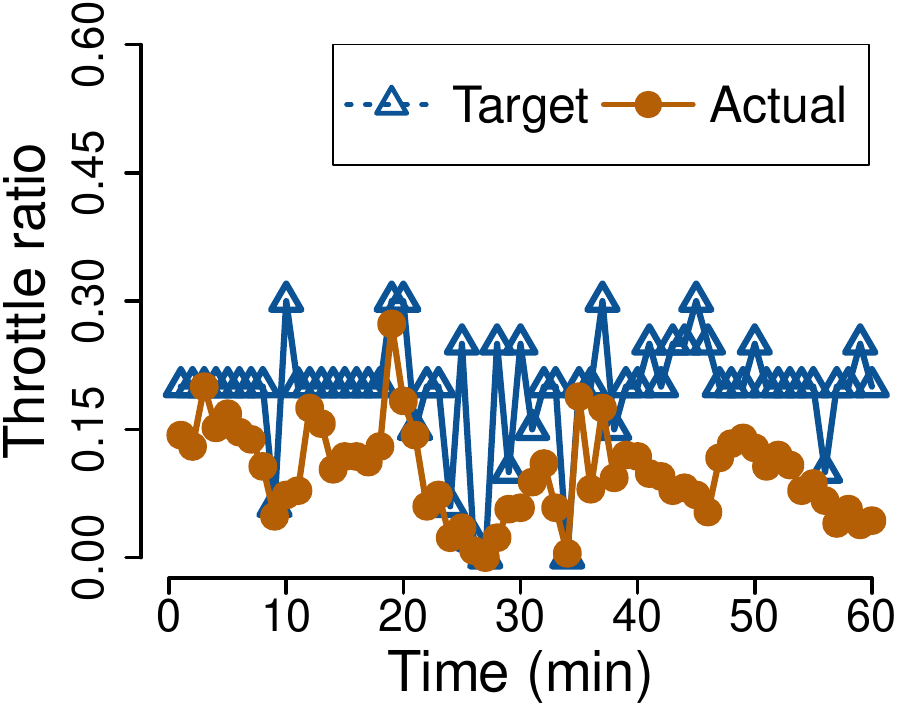}
    \caption{\texttt{media-filter-service}}
    \label{fig:social_diurnal_actual_throttle_mediafilter}
  \end{subfigure}
  \hfill
  \begin{subfigure}{0.496\columnwidth}
    \includegraphics[width=1\columnwidth]{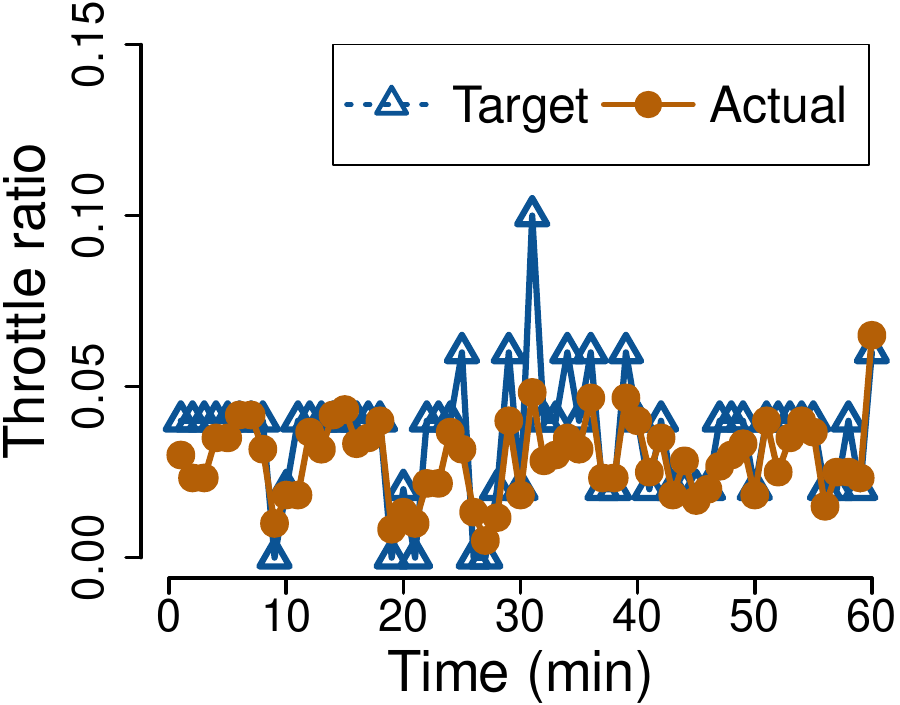}
    \caption{\texttt{post-storage-service}}
    \label{fig:social_diurnal_actual_throttle_poststorage}
  \end{subfigure}
    \caption{{\syslocal} is able to follow the given performance target over
    time, by adjusting per-service resources.}
    \label{fig:social_diurnal_actual_throttle}
  \vspace{100pt}
\end{figure}

We demonstrate {\syslocal}'s ability to achieve {\sysglobal}'s given performance
target of CPU throttle ratio (\S\ref{subsec:local_plane}). This is one factor that determines {\sysname}'s
effectiveness in maintaining the end-to-end SLO. To this end,
Figure~\ref{fig:social_diurnal_actual_throttle} dives into Social-Network and
illustrates two services from ``High'' and ``Low'' CPU usage groups
(\S\ref{sec:appendix_cluster}): \texttt{media-filter-service} and
\texttt{post-storage-service}. Two subfigures compare the target throttle
ratio and the actual throttle ratio, over a period of 60 min. {\syslocal}'s
heuristics meets the targets relatively well and reacts quickly to target
changes, especially when the target is low
(Figure~\ref{fig:social_diurnal_actual_throttle_poststorage}). In
Figure~\ref{fig:social_diurnal_actual_throttle_mediafilter}, we note that the
actual throttle ratio is lower than the target. The reason is that the throttle
ratio is very sensitive to CPU allocation, especially when the target is high.
As a result, {\syslocal} tries to err on the safe side, and it can
over-allocate to avoid exceeding the targeted throttle ratio due to estimation
errors.

\end{document}